\documentclass[12pt]{article}

\makeatletter
 \@addtoreset{equation}{section}
 
\makeatother
\usepackage{amssymb}
\usepackage{amsbsy}
\usepackage{theorem}
\usepackage[dvips]{graphicx}
\usepackage{wick}

\theorembodyfont{\rmfamily}

\begin{document}
%%%%%%%%%%% Title Page %%%%%%%%%%%%%%%%%%%%

\begin{titlepage}

\renewcommand{\thefootnote}{\fnsymbol{footnote}}

\begin{flushright}
\begin{tabular}{l}
UTHEP-600\\
RIKEN-TH-177\\
\end{tabular}
\end{flushright}

\bigskip

\begin{center}
{\Large \bf 
Light-cone Gauge Superstring Field Theory\\ 
and Dimensional Regularization II
\\}
\end{center}

\bigskip

\begin{center}
%% AUTHORS
{\large Yutaka Baba}${}^{a}$\footnote{e-mail:
        ybaba@riken.jp},
{\large Nobuyuki Ishibashi}${}^{b}$\footnote{e-mail:
        ishibash@het.ph.tsukuba.ac.jp},
{\large Koichi Murakami}${}^{a}$\footnote{e-mail:
        murakami@riken.jp}
\end{center}

\begin{center}
${}^{a}${\it 
Theoretical Physics Laboratory, Nishina Center, RIKEN,\\
Wako, Saitama 351-0198, Japan}
\end{center}
\begin{center}
${}^{b}${\it
Institute of Physics, University of Tsukuba,\\
Tsukuba, Ibaraki 305-8571, Japan}\\
\end{center}

\bigskip

\bigskip

\bigskip

\begin{abstract}
We propose a dimensional regularization scheme to deal with 
the divergences caused by colliding supercurrents inserted 
at the interaction points, in the light-cone gauge NSR superstring 
field theory.
We formulate the theory in $d$ dimensions and define the amplitudes 
as analytic functions of $d$. 
With an appropriately chosen three-string interaction term and 
large negative $d$, 
the tree level amplitudes for the (NS,NS) closed strings can be 
recast into a BRST invariant form, using the superconformal field 
theory  proposed in Ref.~\cite{Baba:2009fi}.
We show that in the limit $d\rightarrow 10$ they coincide with the 
results of the first quantized theory.
Therefore we obtain the desired results without adding 
any contact interaction terms to the action.
\end{abstract}

\setcounter{footnote}{0}
\renewcommand{\thefootnote}{\arabic{footnote}}

\end{titlepage}

%%%%%%%%%%%%%%%%%%%%%%%%%%%%%%%%%%%%%

\section{Introduction}

Perturbative expansion of amplitudes in the light-cone gauge 
NSR superstring field theory~\cite{Mandelstam:1974hk,Sin:1988yf} 
involves divergences even at the tree level. 
Transverse supercurrents are inserted at the interaction points 
of the joining-splitting interaction and divergences arise 
when they get close to each other. 
Similar divergences exist in other superstring field theories
~\cite{Greensite:1986gv,Greensite:1987hm,%%%
                 Greensite:1987sm,Green:1987qu,Wendt:1987zh}. 

In the previous paper~\cite{Baba:2009kr} we have proposed 
a dimensional regularization scheme to deal with these divergences. 
In the light-cone gauge, one can define the theory in $d~(d\neq 10)$ dimensions. 
Taking $d$ to be largely negative, we can make 
the tree level amplitudes finite. 
Defining the amplitudes for such $d$, one can obtain the
amplitudes for $d=10$ by analytic continuation. 
Since what matters is the Virasoro central charge on the 
worldsheet, 
one can effectively change $d$ also by using 
conformal field theory 
other than that for the transverse 
variables $X^i,\psi^i,\tilde{\psi}^i$. 
In Ref.~\cite{Baba:2009kr}, we have proposed one such scheme and 
shown that the results of the first quantized formulation can
be reproduced by such a procedure, in the case of the four string 
amplitudes. 

In order for the dimensional regularization scheme to be effective,
it should preserve as many symmetries of the theory as possible. 
In Refs.~\cite{Baba:2009ns,Baba:2009fi},
we have shown that the light-cone gauge string field theory 
in noncritical spacetime dimensions 
corresponds to a BRST invariant worldsheet theory
with the longitudinal variables and the ghosts. 
Since the BRST symmetry on the worldsheet is supposed to 
be related to the gauge symmetry of the string field theory, 
these results imply that  the dimensional regularization 
can be carried out with the gauge symmetry preserved. 

In this paper, we would like to propose a dimensional regularization
scheme for the light-cone gauge NSR superstring field theory,
in which the results of 
Ref.~\cite{Baba:2009fi} can be used. 
We just formulate the theory in $d$ dimensions and 
define the amplitudes as analytic functions of $d$. In this paper, 
we deal with closed string field theory and restrict
ourselves to the amplitudes with only the (NS,NS) external lines. 
We show that the tree level amplitudes can be recast into 
a BRST invariant form using the superconformal field theory proposed 
in Ref.~\cite{Baba:2009fi}.
In this form, it is easy to show that the amplitudes coincide with
the results of the first quantized formulation without any need for
the modification of the action by adding the counterterms. 

The organization of this paper is as follows. 
In section~\ref{sec:dimensional-reg},
we study the light-cone gauge closed string field theory 
for NSR superstrings
defined in spacetime dimension $d\neq 10$.
We show that the tree level amplitudes become
well-defined by setting $d$ to be a sufficiently
large negative value.
In section~\ref{sec:BRST-invariant-form},
we rewrite the tree level amplitudes 
into a BRST invariant form, using the 
superconformal field theory for the longitudinal variables
$X^{\pm},\psi^{\pm},\tilde{\psi}^{\pm}$
formulated in Ref.~\cite{Baba:2009fi}
and introducing the ghost fields.
In section~\ref{sec:counterterms},
we show that the tree level amplitudes coincide with the results 
of the first quantized formulation in the limit $d\to10$. 
Section~\ref{sec:Conclusions-and-Discussions}
is devoted to conclusions and discussions. 
In appendix~\ref{sec:action}, 
we explain the details of the action 
of the superstring field theory given in 
section~\ref{sec:dimensional-reg}.
In appendix~\ref{sec:Amplitudes},
we present the calculations to obtain
the tree level amplitudes. 
In appendix~\ref{sec:psiminus}, we present a proof 
of the property satisfied by the
correlation functions of $\psi^{-}$, which is used
in section~\ref{sec:BRST-invariant-form}.

%%%%%%%%%%%%%%%%%%%%%%%%%%%%%%%%%%%%%%%%%%

\section{Amplitudes for $d\neq 10$ and Dimensional Regularization}
\label{sec:dimensional-reg}

In order to dimensionally regularize the light-cone gauge
NSR string field theory, 
we take the worldsheet theory to be the free theory 
of the transverse variables $X^i,\psi^i,\tilde{\psi}^i$
$(i=1,\cdots ,d-2)$.
%\footnote{This is a bit different from 
%   the prescription in Ref.~\cite{Baba:2009kr}, 
%   in which the worldsheet theory is taken to be that for the 
%   transverse variables $X^i,\psi^i,\tilde{\psi}^i$ $(i=1,\cdots ,8)$ 
%   tensored by a CFT with large negative Virasoro central
%   charge $c$. }
The light-cone gauge string field theory can be defined even for 
$d \neq 10$.
In this paper, we concentrate on the closed strings 
in the (NS,NS) sector and the action is given in the form 
\begin{eqnarray}
S & = & \int dt
   \left[\frac{1}{2}\int d1d2
         \left\langle R\left(1,2\right)
               |\Phi (t) \right\rangle _{1}
         \left(i\frac{\partial}{\partial t}
               - \frac{L_{0}^{\mathrm{LC}(2)}
                        +\tilde{L}_{0}^{\mathrm{LC} (2)}
                        -\frac{d-2}{8}}
                      {\alpha_{2}}
         \right)
         \left| \Phi (t) \right\rangle _{2}
    \right.
\nonumber \\
 &  & \hphantom{\int dt\frac{1}{2}\int d1d2}
\left.
\ensuremath{
   \vphantom{\left(i\frac{\partial}{\partial t}
                   -\frac{L_{0}^{\mathrm{LC}\left(2\right)}
                          +\tilde{L}_{0}^{\mathrm{LC}\left(2\right)}
                          -\frac{d-2}{8}}
                         {\alpha_{2}}\right)
             }}
 {}+\frac{2g}{3}
     \int d1d2d3
     \left\langle V_{3} \left(1,2,3\right)|
                  \Phi (t) \right\rangle _{1}
     \left| \Phi (t) \right\rangle _{2}
     \left| \Phi (t) \right\rangle _{3}
\right].
\label{eq:action}
\end{eqnarray}
In order for the amplitudes of the light-cone gauge 
string field theory to be rewritten into a BRST invariant form, 
the three-string interaction term should be 
taken appropriately. 
Details of the action (\ref{eq:action})
are explained in appendix \ref{sec:action}. 

\begin{figure}[htb]
\begin{center}
	\includegraphics[width=38em]{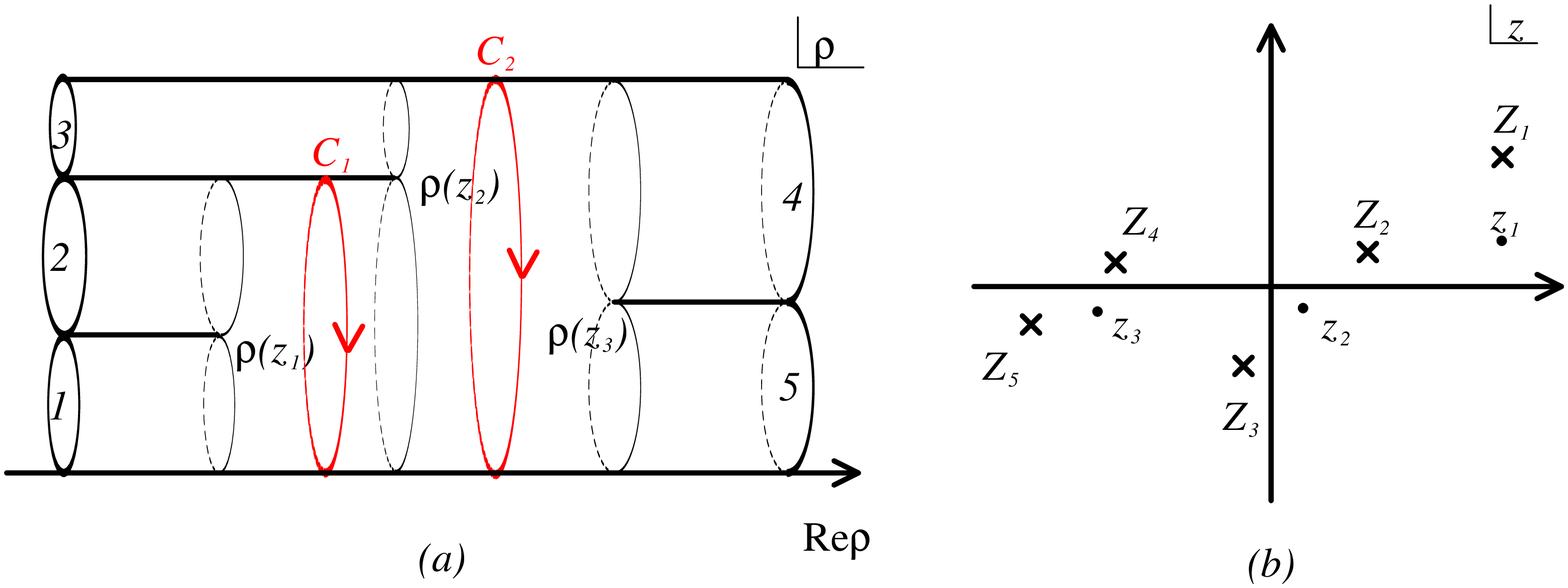}
	\caption{In $(a)$ is depicted 
                 a typical $N$-string tree diagram with $N=5$,
                 on which a complex coordinate
                 $\rho$ is introduced.
                 Via the Mandelstam mapping $\rho (z)$, 
                 the $\rho$-plane 
                 is mapped to the complex $z$-plane in $(b)$.
          $\rho (z_{I})$ $(I=1,2,3)$ are 
          the interaction points on the $\rho$-plane. 
          For the string diagram $(a)$, the complex Schwinger
          parameters $\mathcal{T}_{\mathcal{I}}$
          $(\mathcal{I}=1,2)$ are given by
          $\mathcal{T}_{1}=\rho (z_{2}) - \rho (z_{1})$
           and $\mathcal{T}_{2} = \rho (z_{3}) - \rho (z_{2})$.
          $C_{\mathcal{I}}$ are the contours
          of the integrals in eq.(\ref{eq:BRSTinv})
          for this string diagram.}
	\label{fig:Npt}
\end{center}
\end{figure}

Starting from this action, the tree level $N$-string 
amplitudes can be calculated perturbatively. 
A typical tree level $N$-string diagram is depicted
in Fig.~\ref{fig:Npt}~$(a)$ for the $N=5$ case.
On such string diagrams, we introduce a complex 
$\rho$-coordinate as usual.
The $N$-string tree diagram is mapped to the complex $z$-plane
in Fig.~\ref{fig:Npt}~$(b)$ via the Mandelstam mapping $\rho (z)$
defined as
\begin{equation}
\rho (z) = \sum_{r=1}^{N} \alpha_{r} \ln (z-Z_{r})~,
\label{eq:Mandelstam-N}
\end{equation}
where the external lines are mapped to the regions
$z \sim Z_{r}$ $(r=1,\ldots,N)$.
We denote the interaction points by $z_{I}$
$(1,\ldots, N-2)$ which determined by $\partial \rho (z_{I})=0$.
%%%%%%%%%%%%%%%%%%%%%%%%%%%%%%%%%%%%%%%
%%%%%%%%%%%%%%%%%%%%%%%%%%%%%%%%%%%%%%
The resulting amplitudes can be expressed as 
an integral over the moduli space of the string diagram as
\begin{equation}
\mathcal{A}_{N}
=  \left(4ig\right)^{N-2}
   \int \left(\prod_{\mathcal{I}=1}^{N-3}
              \frac{d^{2}\mathcal{T}_{\mathcal{I}}}{4\pi}\right)
      F_{N}\left(\mathcal{T}_{\mathcal{I}},
                 \bar{\mathcal{T}_{\mathcal{I}}}\right)~,
\label{eq:AN}
\end{equation}
where
$\mathcal{T}_{\mathcal{I}}$ $(\mathcal{I}=1,\ldots,N-3)$
denotes the complex Schwinger parameter
for the $\mathcal{I}$-th internal propagator.
$\mathcal{T}_{\mathcal{I}}$'s  constitute
the $N-3$ complex moduli parameters
of the tree string diagram with $N$ external strings,
and are
the $N$-string generalization of $\mathcal{T}$ 
given in eq.(\ref{eq:Schwinger}) 
for the four-string case.
The integral in eq.(\ref{eq:AN})
is taken over
the whole moduli space of the string diagram.
The integrand 
$F_{N} (\mathcal{T}_{\mathcal{I}},\bar{\mathcal{T}}_{\mathcal{I}})$
is described by using the worldsheet field theory for
the transverse variables~\cite{Baba:2009kr} as
\begin{eqnarray}
F_{N}\left(\mathcal{T}_{\mathcal{I}},
           \bar{\mathcal{T}_{\mathcal{I}}}\right)
 & = & \left(2\pi\right)^{2} 
       \delta \Biggl( \sum_{r=1}^{N}p_{r}^{+} \Biggr)
       \delta \Biggl( \sum_{r=1}^{N}p_{r}^{-} \Biggr)
       \mathrm{sgn}\Biggl(\prod_{r=1}^{N}\alpha_{r}\Biggr)
       e^{-\frac{d-2}{16}
          \Gamma\left[ \ln \left(\partial\rho
                                  \bar{\partial}\bar{\rho}\right)
                \right]}
\nonumber \\
& & \quad \times
    \left\langle 
       \prod_{I=1}^{N-2} \left[
         \left(\partial^{2}\rho\left(z_{I}\right)
               \bar{\partial}^{2} \bar{\rho} \left(\bar{z}_{I}\right)
         \right)^{-\frac{3}{4}}
         T_{F}^{\mathrm{LC}} \left(z_{I}\right)
         \tilde{T}_{F}^{\mathrm{LC}}\left(\bar{z}_{I}\right)
                   \right]
       \prod_{r=1}^{N} V_{r}^{\mathrm{LC}}
    \right\rangle.
\label{eq:FN0}
\end{eqnarray}
Here 
$\langle \mathcal{O} \rangle$ denotes the expectation value 
of the operator $\mathcal{O}$ on the complex $z$-plane,
defined as
\begin{equation}
\left\langle \mathcal{O}\right\rangle  
 =  \frac{\displaystyle \int
          \left[ dX^{i} d\psi^{i} d\tilde{\psi}^{i} \right]
            e^{-S_{\mathrm{LC}}}
            \, \mathcal{O}}
         {\displaystyle \int
          \left[dX^{i} d\psi^{i} d\tilde{\psi}^{i} \right]
          e^{-S_{\mathrm{LC}}}}~,
\label{eq:expectation}
\end{equation}
and $S_{\mathrm{LC}}$ denotes the worldsheet action of 
the light-cone gauge NSR superstring.
$V_{r}^{\mathrm{LC}}$ is the vertex operator defined
in eq.(\ref{eq:VLC}), and $T^{\mathrm{LC}}_{F} (z)$ is
the transverse supercurrent.
$\Gamma\left[ \ln 
\left(\partial\rho\bar{\partial}\bar{\rho}\right)\right]$ 
is given in Ref.~\cite{Baba:2009ns} as\footnote{
We assume $\partial^2 \rho (z_I)\neq 0$ for all $z_I$ which is 
true generically. 
$\partial^2 \rho (z_I)= 0$ when $z_I$ coincides with another interaction point. 
Since such cases are of measure $0$ in the moduli space, 
we treat it as a limit of the generic case, 
in which the interaction points $z_I~(I=1,\cdots ,N-2)$ are all distinct. 
}
\begin{equation}
e^{-\Gamma \left[ \ln 
    \left(\partial\rho\bar{\partial}\bar{\rho}\right)\right]}
 = \left| \sum_{r=1}^{N} \alpha_{r} Z_{r} \right|^{4}
   \prod_{r=1}^{N} 
       \left( |\alpha_{r}|^{-2}
              e^{-2 \mathop{\mathrm{Re}} \bar{N}^{rr}_{00}}
        \right)
   \prod_{I=1}^{N-2} \left| \partial^{2} \rho (z_{I}) \right|^{-1}~,
\label{eq:GammaN}
\end{equation}
where $\bar{N}^{rr}_{00}$ denotes a Neumann coefficient defined as
\begin{equation}
\bar{N}^{rr}_{00} = \frac{\tau^{(r)}_{0}+i\beta_{r}}{\alpha_{r}}
  - \sum_{s\neq r} \frac{\alpha_{s}}{\alpha_{r}} 
    \ln \left( Z_{r} - Z_{s} \right)~,
\qquad
\tau_{0}^{(r)}+i\beta_{r}
  \equiv \rho (z_{I^{(r)}})~.
\label{eq:Nbar-rr-00}
\end{equation}
Here $z_{I^{(r)}}$ denotes the  interaction point on the $z$-plane
at which the $r$-th string interacts.
Which of $z_{I}$ should be identified 
with $z_{I^{(r)}}$ depends on the channel.
For example, 
$z_{I^{(1)}}=z_{I^{(2)}}=z_{1}$,
$z_{I^{(3)}} = z_{2}$
and
$z_{I^{(4)}}=z_{I^{(5)}}=z_{3}$
for the string diagram depicted in Fig.~\ref{fig:Npt}~$(a)$,
while
$z_{I^{(1)}}=z_{1}$, $z_{I^{(2)}}=z_{I^{(3)}}=z_{2}$
and $z_{I^{(3)}}=z_{I^{(4)}}=z_{3}$
for the string diagram in Fig.~\ref{fig:Npt2}.
See appendix~\ref{sec:Amplitudes} for
details of the calculations to obtain 
the expression~(\ref{eq:AN}) of the amplitude.

%%%%%%%%%%%%%%%%%%%%%%%%%%%%%%%%%%%%%%%%%%%%%%%
\begin{figure}[htb]
\begin{center}
  \includegraphics[width=22.5em]{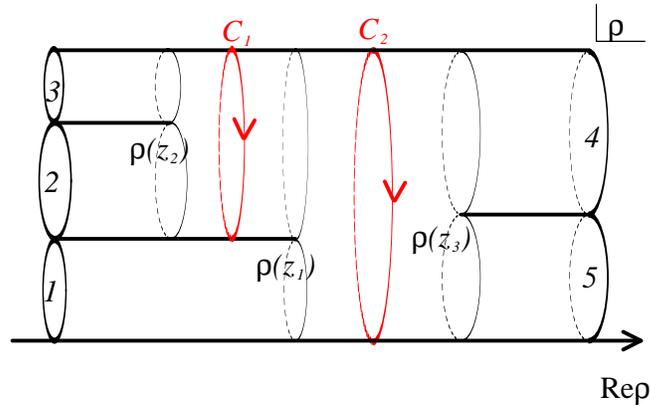}
  \caption{A $5$-string tree diagram
           in a different channel
           from that in Fig.~\ref{fig:Npt}~$(a)$.}
   \label{fig:Npt2}
\end{center}
\end{figure}
%%%%%%%%%%%%%%%%%%%%%%%%%%%%%%%%%%%%%%%%%%%%%%%%%

In general,
$F_{N}\left(\mathcal{T}_{\mathcal{I}},
             \bar{\mathcal{T}_{\mathcal{I}}}\right)$
in eq.(\ref{eq:FN0}) is singular in the limit
$z_{I} \rightarrow z_{J}$.
Nevertheless,
if $d$ is taken to be a sufficiently
large negative value as a regularization,
$F_{N}\left(\mathcal{T}_{\mathcal{I}},
            \bar{\mathcal{T}_{\mathcal{I}}}\right)$
vanishes
in the limit $z_{I} \rightarrow z_{J}$.
It is because in this limit,
$e^{-\frac{d-2}{16} \Gamma 
     \left[ \ln \left( \partial \rho \bar{\partial} \bar{\rho}
                \right) \right]}$
behaves as $\left| z_{I} - z_{J} \right|^{-\frac{d-2}{8}}$
and the contributions of the other operators are 
with $d$ independent power of $\left|z_{I}-z_{J}\right|$. 
The other singularities can be dealt with by the analytic continuation
of the external momenta $p_{r}$.
Thus we can define the integral in eq.(\ref{eq:AN}) for such $d$
and obtain the dimensionally regularized amplitudes.

%%%%%%%%%%%%%%%%%%%%%%%%%%%%%%%%%%%%%%%%%%%%%%%%%%
\section{BRST Invariant Form of Amplitudes}
\label{sec:BRST-invariant-form}

In this section, we would like to show that 
the amplitude (\ref{eq:AN}) can be recast into 
a BRST invariant form
using the superconformal field theory proposed 
in Ref.~\cite{Baba:2009fi}. 
We basically follow the procedure given in 
Refs.~\cite{D'Hoker:1987pr,Aoki:1990yn,Baba:2009kr,Baba:2009ns}. 
In the subsequent calculations, we will not care about 
the overall numerical factor.

We first note that from eq.(\ref{eq:GammaN}) one can obtain
the relation
\begin{eqnarray}
e^{-\frac{d-2}{16}
         \Gamma \left[ \ln \left(\partial\rho
                              \bar{\partial}\bar{\rho} \right)
                \right]}
&=&
e^{-\frac{d-10}{16}
    \Gamma \left[ \ln \left(\partial\rho\bar{\partial}\bar{\rho}
                      \right)
           \right]}
\left| \sum_{s=1}^{N} \alpha_{s} Z_{s} \right|^{2}
\nonumber\\
&& \quad \times
\prod_{r=1}^{N} \left(\left|\alpha_{r}\right|^{-1}
               e^{-\mathop{\mathrm{Re}}\bar{N}_{00}^{rr}}
          \right)
\prod_{I=1}^{N-2} \left(\partial^{2}\rho\left(z_{I}\right)
                \bar{\partial}^{2}\bar{\rho}\left(\bar{z}_{I}\right)
          \right)^{-\frac{1}{4}}.
\label{eq:Gamma-Gamma}
\end{eqnarray}
By using this relation, eq.(\ref{eq:FN0}) becomes
\begin{eqnarray}
\lefteqn{
F_{N}\left(\mathcal{T}_{\mathcal{I}},
           \bar{\mathcal{T}_{\mathcal{I}}}\right)
 \sim  \left(2\pi\right)^{2} 
       \delta \Biggl( \sum_{r=1}^{N}p_{r}^{+} \Biggr)
       \delta \Biggl( \sum_{r=1}^{N}p_{r}^{-} \Biggr)
       e^{-\frac{d-10}{16}
          \Gamma\left[ \ln \left(\partial\rho
                                  \bar{\partial}\bar{\rho}\right)
                \right]}
 \left| \sum_{s=1}^{N} \alpha_{s} Z_{s} \right|^{2}
}\nonumber \\
& & \qquad \times
    \left\langle 
       \prod_{I=1}^{N-2} \left[
         \left(\partial^{2}\rho\left(z_{I}\right)
               \bar{\partial}^{2} \bar{\rho} \left(\bar{z}_{I}\right)
         \right)^{-1}
         T_{F}^{\mathrm{LC}} \left(z_{I}\right)
         \tilde{T}_{F}^{\mathrm{LC}}\left(\bar{z}_{I}\right)
                   \right]
       \prod_{r=1}^{N}\left( \alpha_{r}^{-1}  V_{r}^{\mathrm{LC}}
                             e^{-\mathop{\mathrm{Re}} \bar{N}^{rr}_{00}}
                      \right)
    \right\rangle.
\label{eq:FN1}
\end{eqnarray}

%%%%%%%%%%%%%%%%%%%%%%%%%%%%%%%%%%%%%%%%%%%
\subsection{Ghosts}

In order to obtain a BRST invariant form, 
we need to introduce the longitudinal variables 
and the ghosts. 
Let us first consider the ghost fields $b,c,\beta,\gamma$
and their anti-holomorphic counterparts.
The ghosts can be introduced~\cite{Baba:2009kr} 
by multiplying $F_{N}$ by 
\begin{eqnarray}
&& \int\left[d(\mbox{ghost})\right]
       e^{-S_{\mathrm{gh}}}
       \lim_{z\to\infty}\left(\frac{1}{|z|^{4}}
                             c(z) \tilde{c}(\bar{z})
                        \right)
\nonumber \\
&& \quad \times
   \prod_{I=1}^{N-2} \left[
              b\left(z_{I}\right) \tilde{b}\left(\bar{z}_{I}\right)
              e^{\phi}\left(z_{I}\right)
              e^{\tilde{\phi}}\left(\bar{z}_{I}\right)
            \right]
    \prod_{r=1}^{N} \left[
              c\left(Z_{r}\right)\tilde{c}\left(\bar{Z}_{r}\right)
              e^{-\phi}\left(Z_{r}\right)
              e^{-\tilde{\phi}}\left(\bar{Z}_{r}\right)
           \right],
\end{eqnarray}
which is just a constant. 
Here $S_{\mathrm{gh}}$ denotes the worldsheet action
for the ghost fields.
We have used a shorthand notation
$d(\mbox{ghost})=db dc d\beta d\gamma
 d\tilde{b} d\tilde{c} d\tilde{\beta} d\tilde{\gamma}$,
and bosonized
$\beta\gamma$-ghosts~\cite{Friedan:1985ge} as
\begin{equation}
\beta=e^{- \phi} \partial \xi~,
\qquad
\gamma = \eta e^{\phi}~.
\end{equation}

%%%%%%%%%%%%%%%%%%%%%%%%%%%%%%%%%%%%%%%%%%%%%%%%
\subsection{Longitudinal variables}

%%\subsubsection*{Superconformal field theory for 
%%   longitudinal variables}

Next, let us consider the longitudinal variables
$X^{\pm},\psi^{\pm},\tilde{\psi}^{\pm}$,
which are the component fields of the superfields
$\mathcal{X}^{\pm}$ given as
\begin{equation}
\mathcal{X}^{\pm} (\mathbf{z},\bar{\mathbf{z}})
 = X^{\pm} + i\theta \psi^{\pm} + i\bar{\theta} \tilde{\psi}^{\pm}
   +i \theta \bar{\theta} F^{\pm}~.
\label{eq:superXpm}
\end{equation}
We can rewrite the correlation function on 
the right hand side of eq.(\ref{eq:FN1})
using the $X^\pm$ CFT \cite{Baba:2009fi}. 
The $X^\pm$ CFT is a superconformal field theory 
of the longitudinal variables, with the action 
\begin{equation}
S_{\pm} \equiv 
-\frac{1}{2\pi} \int d^{2} \mathbf{z} \left(
      \bar{D} \mathcal{X}^{+} D \mathcal{X}^{-}
      + \bar{D} \mathcal{X}^{-} D \mathcal{X}^{+} \right)
+\frac{d-10}{8} \Gamma_{\mathrm{super}} \left[\Phi\right]~.
% {}-\frac{1}{2\pi}\int d^{2}z
%        \left(\partial X^{+}\bar{\partial}X^{-}
%              +\partial X^{-}\bar{\partial}X^{+}
%\right. \nonumber \\
%&& \hphantom{-\frac{1}{2\pi}\int d^{2}z}
%    \quad \left.
%       {}+ \psi^{+}\bar{\partial}\psi^{-}
%         + \psi^{-}\bar{\partial}\psi^{+}
%         + \tilde{\psi}^{+} \partial \tilde{\psi}^{-}
%         + \tilde{\psi}^{-} \partial \tilde{\psi}^{+}
%    \right)
%\nonumber \\
%&& {}+\frac{d-10}{8} \Gamma_{\mathrm{super}} \left[\Phi\right]~.
\label{eq:action-pm}
\end{equation}
Here $\Gamma_{\mathrm{super}} \left[\Phi\right]$ is the 
super Liouville action,
%% expressed as
\begin{equation}
\frac{d-10}{8}\Gamma_{\mathrm{super}}\left[\Phi\right] 
 =  -\frac{d-10}{16\pi}\int d^{2}\mathbf{z}\bar{D}\Phi D\Phi\;,
\label{eq:Liouvilleaction}
\end{equation}
and $\Phi$ is the superfield given by 
\begin{eqnarray}
\Phi\left(\mathbf{z},\bar{\mathbf{z}} \right) 
 &=&
   \ln \left( -4
             \left(D\Theta^{+}\right)^{2}\left(\mathbf{z}\right)
             \left(\bar{D}\bar{\Theta}^{+}\right)^{2}
                     \left(\bar{\mathbf{z}}\right)
        \right)\;,
\nonumber\\
\Theta^{+}\left(\mathbf{z}\right) 
 &=&
   \frac{D\mathcal{X}^{+}}
        {(\partial \mathcal{X}^{+})^{\frac{1}{2}}}
    \left(\mathbf{z}\right)\;.
%= \frac{i \psi^{+}}{(\partial X^{+})^{\frac{1}{2}}} (z)
% + \theta \left[ (\partial X^{+})^{\frac{1}{2}} (z)
%                 - \frac{1}{2} 
%                   \frac{\psi^{+} \partial \psi^{+}}
%                        {(\partial X^{+})^{\frac{3}{2}}}  (z)
%           \right].
\label{eq:Phi}
\end{eqnarray}

In this superconformal field theory, we have
\begin{eqnarray}
&&
\int \left[dX^{\pm}d\psi^{\pm} d\tilde{\psi}^{\pm} \right]
    e^{-S_{\pm}}
    \, F\left[ X^{+},\psi^{+},\tilde{\psi}^{+}\right]
    \prod_{r=1}^{N} 
        e^{-ip_{r}^{+}X^{-}} \left(Z_{r},\bar{Z}_{r}\right)
\nonumber \\ 
&& \qquad \sim
    e^{-\frac{d-10}{16} \Gamma\left[ \ln \left( \partial \rho
                                       \bar{\partial} \bar{\rho}
                                         \right) \right]}
  F\left[-\frac{i}{2} \left( \rho+ \bar{\rho} \right),
         0,0 \right]~,
\label{eq:CFTvev}
\end{eqnarray}
for any functional
$F \left[  X^{+},\psi^{+},\tilde{\psi}^{+}\right]$
of $X^{+}$, $\psi^{+}$, $\tilde{\psi}^{+}$.
This can be obtained from eq.(2.11) of Ref.~\cite{Baba:2009fi}
by setting all the Grassmann odd coordinates of the external lines 
$\Theta_{r}$ $(r=1,\ldots,N)$ to be $0$. 
By using eq.(\ref{eq:CFTvev}), we obtain
\begin{eqnarray}
&&\int [dX^{\pm}d\psi^{\pm}d\tilde{\psi}^{\pm}]
e^{-S_{\pm}}\,
   \prod_{r=1}^{N} 
    \left[
     V_{r}^{\prime \mathrm{DDF}} \left(Z_{r},\bar{Z}_{r}\right)
     e^{\frac{d-10}{16} \frac{i}{p^{+}_{r}}X^{+}}
     (z_{I^{(r)}},\bar{z}_{I^{(r)}})
    \right]
\nonumber\\
&& \qquad
   \sim (2\pi)^{2} \delta \Biggl( \sum_{r=1}^{N}p_{r}^{+} \Biggr)
       \delta \Biggl( \sum_{r=1}^{N}p_{r}^{-} \Biggr)
e^{-\frac{d-10}{16} \Gamma\left[ \ln \left( \partial \rho
                                       \bar{\partial} \bar{\rho}
                                         \right) \right]}
 \prod_{r=1}^{N} \left(
\alpha_{r}^{-1} V_{r}^{\mathrm{LC}} 
e^{-\mathop{\mathrm{Re}} \bar{N}^{rr}_{00}}
\right)~.
\label{eq:VDDF-VLC}
\end{eqnarray}
The vertex operator $V^{\prime \mathrm{DDF}}_{r}$
on the left hand side is defined as
\begin{equation}
V_{r}^{\prime\mathrm{DDF}}\left(Z_{r},\bar{Z}_{r}\right)
   \equiv 
    :V_{r}^{\mathrm{DDF}}
      e^{-\frac{d-10}{16} \frac{i}{p_{r}^{+}}X^{+}}
      \left(Z_{r},\bar{Z}_{r}\right):~,
\end{equation}
and $V_{r}^{\mathrm{DDF}}$ is the vertex operator
for the DDF state
which corresponds to $V_{r}^{\mathrm{LC}}$ in eq.(\ref{eq:VLC}), 
defined as
\begin{eqnarray}
\lefteqn{
V_{r}^{\mathrm{DDF}}\left(Z_{r},\bar{Z}_{r}\right) 
}\nonumber \\
&& \equiv 
%%\epsilon^{(r)}_{\{ij\}} 
          A_{-n_{1}}^{i_{1} (r)}
         \cdots
         \tilde{A}_{-\tilde{n}_{1}}^{\tilde{\imath}_{1} (r)}
         \cdots 
         B_{-s_{1}}^{j_{1} (r)}
         \cdots\tilde{B}_{-\tilde{s}_{1}}^{\tilde{\jmath}_{1} (r)}
         \cdots \, 
     e^{ip_{r}^{i}X^{i}-ip_{r}^{+}X^{-}
         -i\left(p_{r}^{-}-\frac{N_{r}}{p_{r}^{+}}\right)X^{+}}
    \left(Z_{r},\bar{Z}_{r}\right).~~~
\label{eq:DDF}
\end{eqnarray}
Here $A^{i(r)}_{-n}$ and $B^{i(r)}_{-s}$ denote
the DDF operators given by
\begin{eqnarray}
A_{-n}^{i (r)}
 & \equiv & \oint_{Z_{r}}\frac{dz}{2\pi i}
              \left( i\partial X^{i}
                     +\frac{n}{p_{r}^{+}}\psi^{i}\psi^{+}
              \right)
            e^{-i\frac{n}{p_{r}^{+}}X_{L}^{+}} (z)\ ,
\nonumber \\
B_{-s}^{i (r)} 
& \equiv & \oint_{Z_{r}}\frac{dz}{2\pi i}
            \left(\psi^{i}
                  -\partial X^{i}\frac{\psi^{+}}{\partial X^{+}}
                  -\frac{1}{2} \psi^{i}
                   \frac{\psi^{+} \partial \psi^{+}}
                        {\left(\partial X^{+} \right)^{2}}
            \right)
            \left( \frac{i\partial X^{+}}{p_{r}^{+}}
            \right)^{\frac{1}{2}}
            e^{-i\frac{s}{p_{r}^{+}}X_{L}^{+}} (z)~,
\end{eqnarray}
and $N_{r}$ is the level number,
\begin{equation}
N_{r} 
\equiv \sum_{i}n_{i}+\sum_{j}s_{j}
 =\sum_{k}\tilde{n}_{k}+\sum_{l}\tilde{s}_{l}~.
\end{equation}
The on-shell condition (\ref{eq:on-shell}) implies that $p_{r}^{-}$
is given as 
\begin{equation}
p_{r}^{-}
 =  \frac{1}{p_{r}^{+}}
    \left(\frac{1}{2} \vec{p}_{r}^{\; 2}
          +N_{r}-\frac{d-2}{16}\right).
\label{eq:on-shell-2}
\end{equation}

We need some care to precisely define the operator
$e^{
   \frac{d-10}{16}
   \frac{i}{p_{r}^{+}}
   X^+ } ( z_{I^{(r)}},\bar{z}_{I^{(r)}} )$ 
in the path integral (\ref{eq:VDDF-VLC}) with the action $S_{\pm}$. 
The argument 
$z_{I^{(r)}}$ itself depends on $\alpha_s ,Z_s$ 
and it is influenced by the presence of other operators.%%%
%%%%%%%%%%%%
\footnote{
  In Ref.~\cite{Baba:2009ns}, we were not precise enough 
  about this point. The operator 
  $e^{ \frac{d-26}{24}
       \frac{i}{p_r^+}
       X^+}
   ( z_{I^{(r)}},\bar{z}_{I^{(r)}} )$ 
which appears in eq.(4.2) of Ref.~\cite{Baba:2009ns} 
should have been defined as
$$
\oint_{z_{I^{(r)}}}\frac{dz}{2\pi i}\partial \ln \partial X^+ (z)
\oint_{\bar{z}_{I^{(r)}}}\frac{d\bar{z}}{2\pi i}
\bar{\partial} \ln \bar{\partial} X^+ (\bar{z})
e^{
\frac{d-26}{24}
\frac{i}{p_r^+}
X^+}
(z,\bar{z})~.
$$
}
%%%%%%%%%%%%%%%%%%% 
Here we take the expression
\begin{equation}
\oint_{z_{I^{(r)}}} 
       \frac{d\mathbf{z}}{2\pi i} D\Phi (\mathbf{z})
     \oint_{\bar{z}_{I^{(r)}}}
       \frac{d\bar{\mathbf{z}}}{2\pi i} \bar{D}
          \Phi (\bar{\mathbf{z}})
       e^{\frac{d-10}{16} \frac{i}{p^{+}_{r}} \mathcal{X}^{+}}
        (\mathbf{z},\bar{\mathbf{z}})~.
\label{eq:insertion}
\end{equation}
As the definition of this operator,
this coincides with 
$e^{
\frac{d-10}{16}
\frac{i}{p_{r}^{+}} X^+}
 ( z_{I^{(r)}}, \bar{z}_{I^{(r)}} )$ 
under the identification 
$X^+ \sim -\frac{i}{2} \left( \rho + \bar{\rho} \right)$;
$\psi^+,\tilde{\psi}^{+} \sim 0$,
which can be done
in the path integral of the form on the left hand side
of eq.(\ref{eq:CFTvev}).

%%%%%%%%%%%%%%%%%%%%%%%%%%%

We can introduce the longitudinal variables by substituting 
eqs.(\ref{eq:VDDF-VLC}) and (\ref{eq:insertion})
into eq.(\ref{eq:FN1}). 
With the ghost variables introduced above, 
and using the relation 
$\frac{b}{\partial^{2} \rho} (z_{I})
  = \oint_{z_{I}} \frac{dz}{2\pi i} \frac{b}{\partial \rho} (z)$, 
we obtain
\begin{eqnarray}
F_{N} &\sim& 
  \int \left[dXd\psi d\tilde{\psi} d(\mbox{ghost})\right]
  e^{-S}
  \lim_{z\to\infty} \left(\frac{1}{|z|^{4}}c(z)\tilde{c}(\bar{z})
                    \right)
  \left|\sum_{r=1}^{N} \alpha_{r}Z_{r}\right|^{2}
\nonumber \\
&&  %%\hphantom{dXd\psi d\left(\mathrm{ghost}\right)}
  \quad \times \prod_{r=1}^{N} 
   \left[
     ce^{-\phi} \tilde{c} e^{-\tilde{\phi}}
     V_{r}^{\prime \mathrm{DDF}}\left(Z_{r},\bar{Z}_{r}\right)
         \oint_{z_{I^{(r)}}} 
           \frac{d\mathbf{z}}{2\pi i} D\Phi (\mathbf{z})
         \oint_{\bar{z}_{I^{(r)}}}
           \frac{d\bar{\mathbf{z}}}{2\pi i} \bar{D}
               \Phi (\bar{\mathbf{z}})
          e^{\frac{d-10}{16} \frac{i}{p^{+}_{r}} \mathcal{X}^{+}}
        (\mathbf{z},\bar{\mathbf{z}})
\right]
\nonumber \\
&&  %%\hphantom{dXd\psi d\left(\mathrm{ghost}\right)}
   \quad
    \times \prod_{I=1}^{N-2}
      \left[\oint_{z_{I}}\frac{dz}{2\pi i}
              \frac{b}{\partial\rho} (z)
             e^{\phi} T^{\mathrm{LC}}_{F} (z_{I}) 
            \oint_{\bar{z}_{I}}\frac{d\bar{z}}{2\pi i}
               \frac{\tilde{b}}{\bar{\partial}\bar{\rho}} (\bar{z})
               e^{\tilde{\phi}}
               \tilde{T}^{\mathrm{LC}}_{F} (\bar{z}_{I})
    \right],
\label{eq:FN2}
\end{eqnarray}
up to an overall constant factor,
where
\begin{equation}
S  =  S_{\pm} + S_{\mathrm{LC}}
                     +S_{\mathrm{gh}}~.
\label{eq:actionS}
\end{equation}

%%%%%%%%%%%%%%%%%%%%%%%%
\subsection{$e^{\phi}T_{F}^{\mathrm{LC}}\left(z_{I}\right)$ 
             and the picture changing operator}
\label{sec:picture}

$F_{N}$ is now expressed by the worldsheet theory 
with the action $S$ 
given in eq.(\ref{eq:actionS}).
As was shown in Ref.~\cite{Baba:2009fi}, 
this system possesses a nilpotent BRST charge,
which can be written using the superfields as
\begin{equation}
Q_{\mathrm{B}} 
=  \oint\frac{d\mathbf{z}}{2\pi i}
    \left[-C \left( T_{X^{\pm}}
                    +T_{\mathrm{LC}} \right)
         -C(DC)(DB)
         + \frac{3}{4} (DC)^{2}B
\right]
+ \mbox{c.c.}~,
%= \oint \frac{dz}{2\pi i} \left[
%     c T_{B} + c \left( -\frac{1}{2} (\partial \phi)^{2}
%                        - \partial^{2} \phi - \eta \partial \xi
%                 \right)
%    + c \partial c b
%    - \eta e^{\phi} T_{F}
%    -\frac{1}{4} b \eta \partial \eta e^{2\phi} \right]
%+ \mbox{c.c.}~,
\end{equation}
where $C(\mathbf{z})$ and $B(\mathbf{z})$ are 
the ghost and the anti-ghost superfields,
$T_{\mathrm{LC}} (\mathbf{z})
  =T_{F}^{\mathrm{LC}} + \theta T_{B}^{\mathrm{LC}}$
denotes the transverse super energy-momentum tensor,
and $T_{X^{\pm}} (\mathbf{z})$ is the
super energy-momentum tensor of the $X^{\pm}$ CFT
defined as
\begin{eqnarray}
T_{X^{\pm}} (\mathbf{z})
 &\equiv& \frac{1}{2} D\mathcal{X}^{+} \partial \mathcal{X}^{-}
   +\frac{1}{2} D\mathcal{X}^{-} \partial \mathcal{X}^{+}
   - \frac{d-10}{4}
      S \left( \mathbf{z},
               \mbox{\boldmath$\mathcal{X}$}^{+}_{\! L} \right)~,
\nonumber\\
S \left( \mathbf{z},\mbox{\boldmath$\mathcal{X}$}^{+}_{\! L} 
  \right)
 &\equiv&
  \frac{D^{4}\Theta^{+}}{D\Theta^{+}}
   - 2 \frac{D^{3}\Theta^{+} D^{2} \Theta^{+}}
            {(D\Theta^{+})^{2}}~.
\end{eqnarray}
{}From $Q_{\mathrm{B}}$, the picture changing operator $X$
is obtained as
\begin{equation}
X (z)  \equiv  \left\{ Q_{\mathrm{B}} \, , \, \xi (z)\right\}
 =  c \partial \xi (z)  -e^{\phi}T_{F} (z)
       + \frac{1}{4} \partial b\eta e^{2\phi} (z)
       + \frac{1}{4} b \left( 2 \partial \eta e^{2 \phi}
                              + \eta\partial e^{2\phi} \right) (z)~,
\label{eq:XQBxi}
\end{equation}
where $T_{F}$ is the supercurrent of the matter sector,
namely the lower component of $T_{X^{\pm}} + T_{\mathrm{LC}}$,
given by
\begin{eqnarray}
 T_{F} (z)
  &\equiv&  T_{F}^{\mathrm{LC}} (z)
        + \frac{i}{2} \left( \partial X^{+} \psi^{-}
                            +\partial X^{-} \psi^{+} 
                      \right) (z)
\nonumber\\
&&   {}-\frac{d-10}{4}i\left[
    \left( \frac{5\left(\partial^{2}X^{+}\right)^{2}}
                {4\left(\partial X^{+}\right)^{3}}
           -\frac{\partial^{3}X^{+}}{2\left(\partial X^{+}\right)^{2}}
    \right)\psi^{+}     
    \right.
\nonumber\\
&& 
   \hphantom{-\frac{d-10}{4}i \quad}
   \left.
   \vphantom{
             \left(\frac{5\left(\partial^{2}X^{+}\right)^{2}}
                        {4\left(\partial X^{+}\right)^{3}}\right)
             }
   {}- \frac{2\partial^{2}X^{+}}
            {\left(\partial X^{+}\right)^{2}}
       \partial\psi^{+}  
     + \frac{\partial^{2}\psi^{+}}{\partial X^{+}}
     - \frac{\psi^{+}\partial\psi^{+}\partial^{2}\psi^{+}}
            {2\left(\partial X^{+}\right)^{3}}
     \right] (z)~.
\end{eqnarray}
In the correlation functions of the $X^{\pm}$ CFT
with the insertion 
$\prod_{r=1}^{N} e^{-ip^{+}_{r} X^{-}}(Z_{r},\bar{Z}_{r})$,
the variables
$X^{-}$, $\psi^{-}$, $\tilde{\psi}^{-}$ 
may have poles at the interaction points $z_{I}$,
even if no operators are there.
%~\cite{Baba:2009fi}.
However, the supercurrent $T_{F}$ and thus
the picture changing operator $X$ are regular
at $z_{I}$, when no operators are inserted
there~\cite{Baba:2009fi}.

As a final step to recast the amplitude $\mathcal{A}_{N}$
in eq.(\ref{eq:AN}) into a BRST invariant form,
in the following
we will show that the insertion $e^{\phi}T^{\mathrm{LC}}_{F} (z_{I})$
in the path integral (\ref{eq:FN2})
can be replaced by the picture changing operator
$X(z_{I})$ and thus
\begin{eqnarray}
F_{N} &\sim& 
  \int \left[dXd\psi d\tilde{\psi} d(\mbox{ghost})\right]
  e^{-S}
  \lim_{z\to\infty} \left(\frac{1}{|z|^{4}}c(z)\tilde{c}(\bar{z})
                    \right)
  \left|\sum_{r=1}^{N} \alpha_{r}Z_{r}\right|^{2}
\nonumber \\
&&  %%\hphantom{dXd\psi d\left(\mathrm{ghost}\right)}
  \quad \times \prod_{r=1}^{N} 
   \left[
     ce^{-\phi} \tilde{c} e^{-\tilde{\phi}}
     V_{r}^{\prime \mathrm{DDF}}\left(Z_{r},\bar{Z}_{r}\right)
        \oint_{z_{I^{(r)}}} 
           \frac{d\mathbf{z}}{2\pi i} D\Phi (\mathbf{z})
        \oint_{\bar{z}_{I^{(r)}}}
           \frac{d\bar{\mathbf{z}}}{2\pi i} \bar{D}
              \Phi (\bar{\mathbf{z}})
           e^{\frac{d-10}{16} \frac{i}{p^{+}_{r}}
               \mathcal{X}^{+}}
            (\mathbf{z},\bar{\mathbf{z}})
    \right]
\nonumber \\
&&  %%\hphantom{dXd\psi d\left(\mathrm{ghost}\right)}
   \quad 
    \times \prod_{I=1}^{N-2}
      \left[\oint_{z_{I}}\frac{dz}{2\pi i}
              \frac{b}{\partial\rho} (z)
             X (z_{I}) 
            \oint_{\bar{z}_{I}}\frac{d\bar{z}}{2\pi i}
               \frac{\tilde{b}}{\bar{\partial}\bar{\rho}} (\bar{z})
               \tilde{X} (\bar{z}_{I})
    \right].~~~~~
\label{eq:FN3}
\end{eqnarray}
We would like to show this by proving that the right hand side 
is equal to that of eq.(\ref{eq:FN2}).

Let us introduce a nilpotent fermionic charge $Q$ \cite{Baba:2009kr} as 
\begin{equation}
Q \equiv 
  \oint\frac{dz}{2\pi i} \partial\rho
    \left[c\left(i\partial X^{+}-\frac{1}{2}\partial\rho\right)
          +\frac{1}{2}\eta e^{\phi} \psi^{+}\right](z)~.
\end{equation}
One can show 
\begin{eqnarray}
\lefteqn{
\oint_{z_{I}} \frac{dz}{2\pi i} \frac{b}{\partial \rho} (z)
  X(z_{I})
%= \oint_{z_{I},w} \frac{dz}{2\pi i} \frac{b}{\partial \rho} (z)
%    \oint_{z_{I}} \frac{dw}{2\pi i} \frac{X(w)}{w-z_{I}}
}
\nonumber\\
&=& - \oint_{z_{I}} \frac{dz}{2\pi i} \frac{b}{\partial \rho} (z)
      e^{\phi} T^{\mathrm{LC}}_{F} (z_{I}) 
  + \left[ Q \,,\,
           \oint_{z_{I},w} \frac{dz}{2\pi i}
             \frac{b}{\partial \rho}(z)
           \oint_{z_{I}} \frac{dw}{2\pi i} 
              \frac{\mathcal{O} (w) e^{\phi}(z_{I})}{w-z_{I}}
     \right]
\nonumber\\
&&{}+ \oint_{z_{I},w} \frac{dz}{2\pi i} \frac{b}{\partial \rho}(z)
    \oint_{z_{I}} \frac{dw}{2\pi i} \frac{1}{w-z_{I}}
      \frac{1}{2}
      \left(1 - \frac{i\partial^{2}X^{+}}{\partial^{2}\rho} (w)
      \right) \partial \rho \psi^{-} (w) e^{\phi}(z_{I})~,
\label{eq:X-TF}
\end{eqnarray}
where 
\begin{eqnarray}
\mathcal{O} 
 &\equiv&
    \frac{i}{\partial \rho} \partial X^{-} e^{-\phi} \partial \xi
      + \frac{1}{2\partial^{2} \rho} \partial b \psi^{-}
\nonumber\\
  && {}- \frac{d-10}{4} i \left[
      \left( \frac{5(\partial^{2} X^{+})^{2}}
                  {4 (\partial X^{+})^{3}}
             - \frac{\partial^{3} X^{+}}{(\partial X^{+})^{2}}
       \right)
        \frac{2 e^{-\phi} \partial \xi}{\partial \rho}
      - \frac{2 \partial^{2} X^{+}}{(\partial X^{+})^{2}}
        \partial \left( \frac{2 e^{-\phi} \partial \xi}
                             {\partial \rho}
                 \right)
       \right.
\nonumber\\
&& \hphantom{- \frac{d-10}{4} i \left[\right.} \quad
    \left.
    {}+ \frac{1}{\partial X^{+}}
         \partial^{2} \left( \frac{2 e^{-\phi} \partial \xi}
                                  {\partial \rho} \right)
      - \frac{2 e^{-\phi}\partial \xi}{\partial \rho}
         \frac{\partial \psi^{+} \partial^{2}\psi^{+}}
              {2 (\partial X^{+})^{3}}
       \right]~.
\label{eq:rewriteX}
\end{eqnarray}
Using the relations \begin{eqnarray}
 &  & \left\{ Q\,,\,\oint_{z_{I}}\frac{dz}{2\pi i}\frac{b}{\partial\rho}\right\} =\left[Q\,,\, e^{\phi}T_{F}^{\mathrm{LC}}\left(z_{I}\right)\right]=0~,\nonumber \\
 &  & \left[Q\,,\,\oint_{z_{I}}\frac{dw}{2\pi i}\frac{1}{w-z_{I}}\left(1-\frac{i\partial^{2}X^{+}}{\partial^{2}\rho}(w)\right)\partial\rho\psi^{-}(w)e^{\phi}(z_{I})\right]=0~,\end{eqnarray}
 one can easily find that $Q$ (anti)commutes with all the insertions
in the path integral (\ref{eq:FN3}). The second term on the right
hand side of eq.(\ref{eq:X-TF}), which is $Q$-exact, is therefore
irrelevant in the path integral (\ref{eq:FN3}).

Hence the right hand side of eq.(\ref{eq:FN3}) becomes 
\begin{eqnarray}
 &  & \int\left[dXd\psi d\tilde{\psi}d(\mbox{ghost})\right]
   e^{-S}  \lim_{z\to\infty}\left(\frac{1}{|z|^{4}}
                                  c(z)\tilde{c}(\bar{z})\right)
       \left|\sum_{r=1}^{N}\alpha_{r}Z_{r}\right|^{2}
\nonumber \\
 &  & \quad\times\prod_{r=1}^{N}
        \left[
         ce^{-\phi}\tilde{c}e^{-\tilde{\phi}}
          V_{r}^{\prime\mathrm{DDF}}\left(Z_{r},\bar{Z}_{r}\right)
         \oint_{z_{I^{(r)}}}\frac{d\mathbf{z}}{2\pi i}
             D\Phi(\mathbf{z})
         \oint_{\bar{z}_{I^{(r)}}}\frac{d\bar{\mathbf{z}}}{2\pi i}
             \bar{D}\Phi(\bar{\mathbf{z}})
          e^{\frac{d-10}{16}\frac{i}{p_{r}^{+}}\mathcal{X}^{+}}
             (\mathbf{z},\bar{\mathbf{z}})
        \right]
\nonumber \\
 &  & \quad\times\prod_{I=1}^{N-2}
        \left[
          \oint_{z_{I}}\frac{dz}{2\pi i}
             \frac{b}{\partial\rho}(z)e^{\phi}
                 \left[T_{F}^{\mathrm{LC}}+R\right](z_{I})
          \oint_{\bar{z}_{I}}\frac{d\bar{z}}{2\pi i}
              \frac{\tilde{b}}{\bar{\partial}\bar{\rho}}(\bar{z})
                 e^{\tilde{\phi}}
                 \left[\tilde{T}_{F}^{\mathrm{LC}}+\tilde{R}\right]
                     (\bar{z}_{I})
          \right],
\label{eq:FN4}
\end{eqnarray}
 where 
\begin{equation}
R\left(z_{I}\right)
  \equiv \oint_{z_{I}}\frac{dw}{2\pi i} \frac{1}{w-z_{I}}
    \frac{1}{2}
    \left(1-\frac{i\partial^{2}X^{+}}{\partial^{2}\rho}(w)\right)
     \partial\rho\psi^{-}(w)\ .
\label{eq:Rz}
\end{equation}
 Since $\partial\rho\left(z_{I}\right)=0$, the contour integral on
the right hand side of eq.(\ref{eq:Rz}) is nonvanishing only when
$\psi^{-}\left(w\right)$ is singular at $w=z_{I}$. 
By examining the singularities of the correlation functions of 
$\psi^{-}$ carefully,
one can show that 
$R\left(z_{I}\right)$ and $\tilde{R}\left(\bar{z}_{I}\right)$
do not contribute to the correlation function. Since the proof is
rather long, we present it in appendix~\ref{sec:psiminus}. 
Using this fact, the right hand side of eq.(\ref{eq:FN3}) 
coincides with that of eq.(\ref{eq:FN2}) 
and eq.(\ref{eq:FN3}) is proved.

Thus 
the amplitude $\mathcal{A}_{N}$ is given
by substituting eq.(\ref{eq:FN3}) into eq.(\ref{eq:AN}).
By deforming the contours of the integrals
$\oint_{z_{I}}\frac{dz}{2\pi i}\frac{b}{\partial\rho} (z)$,
we eventually obtain the supersymmetrized version
of the expression in Ref.~\cite{Baba:2009ns}:
\begin{eqnarray}
\mathcal{A}_{N} 
  &\sim& \int\left[dXd\psi d\tilde{\psi} d(\mbox{ghost})\right]
      e^{-S} 
\nonumber\\
&& \quad \times
   \int \prod_{\mathcal{I}=1}^{N-3} d^{2} \mathcal{T}_{\mathcal{I}}
   \left( \prod_{\mathcal{I}=1}^{N-3}
          \left[
             \oint_{C_{\mathcal{I}}} \frac{dz}{2\pi i} 
              \frac{b}{\partial \rho} (z)
             \oint_{C_{\mathcal{I}}} \frac{d\bar{z}}{2\pi i}
              \frac{\tilde{b}}{\bar{\partial} \bar{\rho}} (\bar{z})
          \right]
      \prod_{r=1}^{N} \left[
          c \tilde{c} e^{-\phi-\tilde{\phi}}
          V_{r}^{\prime \mathrm{DDF}}\left(Z_{r},\bar{Z}_{r}\right)
                      \right]
   \right.
\nonumber \\
&& \hphantom{\times
             \int \prod_{I=1}^{N-3} d^{2} \mathcal{T}_{I}
             \qquad}
   \times 
      \prod_{r=1}^{N} 
       \oint_{z_{I^{(r)}}} 
       \frac{d\mathbf{z}}{2\pi i} D\Phi (\mathbf{z})
       \oint_{\bar{z}_{I^{(r)}}}
       \frac{d\bar{\mathbf{z}}}{2\pi i} \bar{D}
          \Phi (\bar{\mathbf{z}})
       e^{\frac{d-10}{16} \frac{i}{p^{+}_{r}} \mathcal{X}^{+}}
          (\mathbf{z},\bar{\mathbf{z}})
\nonumber \\
&& \hphantom{ \times
             \int \prod_{I=1}^{N-3} d^{2} \mathcal{T}_{I}
             \qquad}
   \left.  \times
   \prod_{I=1}^{N-2} \left[ X (z_{I}) \tilde{X} (\bar{z}_{I})
                     \right]
\right),
\label{eq:BRSTinv}
\end{eqnarray}
where the integration contour $C_{\mathcal{I}}$ lies around 
the $\mathcal{I}$-th internal propagator $(\mathcal{I}=1,\ldots,N-3)$
of the light-cone diagram for $N$ strings 
as depicted in Fig.~\ref{fig:Npt}~$(a)$.

%%%%%%%%%%%%%%%%%%%%%%%%%%%%%%%%%%%%%
\subsection{BRST invariance}
In the following, we will show 
the BRST invariance of 
the form of the amplitude in eq.(\ref{eq:BRSTinv}).

First, we show that all the insertions other than
$\prod_{\mathcal{I}=1}^{N-3}
          \left[
             \oint_{C_{\mathcal{I}}} \frac{dz}{2\pi i} 
              \frac{b}{\partial \rho} (z)
             \oint_{C_{\mathcal{I}}} \frac{d\bar{z}}{2\pi i}
              \frac{\tilde{b}}{\bar{\partial} \bar{\rho}} (\bar{z})
          \right]$
in the path integral (\ref{eq:BRSTinv}) are BRST invariant.
By using the fact that the superfields
$\Theta^+\left(\mathbf{z}\right)$
and
$e^{ \frac{d-10}{16}
     \frac{i}{p_{r}^{+}}\mathcal{X}^+}
 (\mathbf{z},\bar{\mathbf{z}})$
are primary fields of weight $0$, 
one can easily show that the OPE between
$T_{X^{\pm}} (\mathbf{z})$ and
the operator (\ref{eq:insertion}) is regular. 
Therefore the operator (\ref{eq:insertion}) is BRST invariant. 
$V_{r}^{\prime\mathrm{DDF}}$
can be considered as the vertex operator (\ref{eq:DDF}) for
the DDF state with modified momentum 
\begin{equation}
p_{r}^{\prime-}  
=  p_{r}^{-}+\frac{d-10}{16}\frac{1}{p_{r}^{+}}~,
\end{equation}
and it is a primary field of weight $(\frac{1}{2},\frac{1}{2})$.
Hence $c\tilde{c}e^{-\phi-\tilde{\phi}}
              V_{r}^{\prime\mathrm{DDF}}
            \left(Z_{r},\bar{Z}_{r}\right)$
is BRST invariant.
Finally, because of eq.(\ref{eq:XQBxi}), it is obvious that
$X(z)$ is BRST invariant.

Next, we consider the remaining insertion 
$\oint_{C_{\mathcal{I}}} \frac{dz}{2\pi i} 
  \frac{b}{\partial \rho} (z)$.
It satisfies the relation,
\begin{equation}
\left\{ Q_{\mathrm{B}} \, , \,
       \oint_{C_{\mathcal{I}}} \frac{dz}{2\pi i} 
         \frac{b}{\partial \rho} (z)
\right\}
= \oint_{C_{\mathcal{I}}} \frac{dz}{2\pi i} 
         \frac{T^{\mathrm{total}}_{B}}{\partial \rho} (z)~,
\label{eq:quasiconformal}
\end{equation}
where $T^{\mathrm{total}}_{B}(z)$ 
is the energy-momentum tensor of the total system.
%$T^{\mathrm{gh}}_{B}(z)$ denotes that of the ghost sector,
%and $T_{B}(z)$ is defined as
%\begin{equation}
% T_{B}(z) \equiv T_{B}^{\mathrm{LC}}(z)
%         + \partial X^{+} \partial X^{-} (z)
%         + T_{B}^{\mathrm{Liouville}} (z)~.
%\end{equation}
%Here $T^{\mathrm{LC}}_{B}(z)$ is the transverse 
%energy-momentum tensor
%and $T_{B}^{\mathrm{Liouville}} (z)$ is defined as
%\begin{eqnarray}
%T_{B}^{\mathrm{Liouville}} (z)
%&\equiv&
% -\frac{d-10}{4}
%      \left[ \frac{1}{2}
%             \left( 
%               \frac{\partial^{3}X^{+}}{\partial X^{+}}
%               - \frac{3}{2}
%                 \frac{(\partial^{2} X^{+})^{2}}
%                      {(\partial X^{+})^{2}}
%            \right)
%      \right.
%\nonumber\\
%&& \hphantom{-\frac{d-10}{4}[}
%      {}-\left( \frac{15}{4} 
%                    \frac{(\partial^{2} X^{+})^{2}}
%                         {(\partial X^{+})^{4}}
%             - \frac{1}{2}
%              \frac{\partial^{3} X^{+}}{(\partial X^{+})^{3}}
%              \right)
%              \psi^{+} \partial \psi^{+}
%       +3 \frac{\partial^{2}X^{+}}{(\partial X^{+})^{3}}
%         \psi^{+} \partial^{2} \psi
%\nonumber\\
%&& \hphantom{-\frac{d-10}{4}[}
%  \left.
%      - \frac{1}{(\partial X^{+})^{2}}
%        \left( \frac{3}{2} 
%          \partial \psi^{+} \partial^{2} \psi^{+}
%          + \frac{1}{2}
%          \psi^{+} \partial^{3} \psi^{+} 
%        \right)
%   \right](z)~.
%\end{eqnarray}
Since the insertion (\ref{eq:quasiconformal})
yields the total derivative with respect to
$\mathcal{T}_{\mathcal{I}}$, the amplitude $\mathcal{A}_{N}$
in eq.(\ref{eq:BRSTinv}) turns out to be BRST invariant
if the surface terms vanish.
The surface terms
correspond to the limits 
$z_{I}\to z_{J}$ and $Z_{r}\to Z_{s}$.
We note that $Z_{r}\to z_{I}$ only when $Z_{r}\to Z_{s}$ 
for some $s$. 
By setting $d$ to be a sufficiently large negative value,
we can make the surface terms
corresponding to the limit $z_{I} \to z_{J}$ vanishing,
as explained in section~\ref{sec:dimensional-reg}.
The limit $Z_{r}\to Z_{s}$ can be dealt with by choosing 
the external momenta appropriately. 
Therefore, with large negative $d$ and 
appropriately chosen external momenta $p_{r}^{\mu}$,
the surface terms are vanishing. 
BRST invariant amplitudes can be defined
by analytically continuing $p_{r}^{\mu}$.

%%%%%%%%%%%%%%%%%%%%%%%%%%%%%%%%%%%
\section{Amplitudes for $d=10$}
\label{sec:counterterms}

Using the BRST invariant form thus obtained, 
let us examine if we can obtain the results of 
the first quantized formalism 
in the limit $d\to 10$. 
Using the standard argument~\cite{Friedan:1985ge}, 
one can change the positions of the picture changing operators 
$X (z)$.
By moving them to $Z_{r}$ $(r=3,\ldots,N)$ and then
deforming the contours of the integrals
$\oint_{C_{\mathcal{I}}}
   \frac{dz}{2\pi i}\frac{b}{\partial \rho} (z)$
as in Ref.~\cite{Baba:2009ns}, we obtain the expression
\begin{eqnarray}
&& \mathcal{A}_{N} \sim
  \int\left[dXd\psi d\tilde{\psi} d(\mbox{ghost})\right]
   e^{-S}    \prod_{s=1,2} 
     \left[ c\tilde{c} e^{-\phi-\tilde{\phi}}
                 V_{s}^{\prime \mathrm{DDF}}
                   \left(Z_{s},\bar{Z}_{s}\right)
     \right]
     c\tilde{c} V_{3}^{\prime (0) \mathrm{DDF}} 
                (Z_{3},\bar{Z}_{3})
\nonumber\\
&& \hphantom{\mathcal{A}_{N} \sim}
  \quad \times
  \int \prod_{s=4}^{N} d^{2} Z_{s}
  \left( \prod_{r=4}^{N} V_{r}^{\prime (0) \mathrm{DDF}}
                         (Z_{r},\bar{Z}_{r})
  \right.
\nonumber\\
&& \hphantom{\mathcal{A}_{N} \sim
             \quad  \int  d^{2} Z_{s} (~\qquad  }
  \left. \times
  \prod_{r=1}^{N} 
     \oint_{z_{I^{(r)}}} 
       \frac{d\mathbf{z}}{2\pi i} D\Phi (\mathbf{z})
       \oint_{\bar{z}_{I^{(r)}}}
       \frac{d\bar{\mathbf{z}}}{2\pi i} \bar{D}
          \Phi (\bar{\mathbf{z}})
       e^{\frac{d-10}{16} \frac{i}{p^{+}_{r}}
          \mathcal{X}^{+}}
         (\mathbf{z},\bar{\mathbf{z}})
    \right),
\label{eq:AN-BRSTinv}
\end{eqnarray}
%Here we have taken the $\beta\gamma$-ghost number anomaly
%into account, and 
where the vertex operator
$V_{r}^{\prime (0) \mathrm{DDF}} (Z_{r},\bar{Z}_{r})$
is defined as
\begin{equation}
V_{r}^{\prime (0) \mathrm{DDF}} (Z_{r},\bar{Z}_{r})
\equiv \left\{ G_{-\frac{1}{2}}\,,\,
    \left[ \tilde{G}_{-\frac{1}{2}} \,,\,
           V^{\prime \mathrm{DDF}}_{r} (Z_{r},\bar{Z}_{r})
    \right] \right\}~,
\end{equation}
and
\begin{equation}
G_{-\frac{1}{2}} \equiv \oint \frac{dz}{2\pi i} T_{F} (z)~,
\qquad
\tilde{G}_{-\frac{1}{2}} \equiv
 \oint \frac{d\bar{z}}{2\pi i} \tilde{T}_{F} (\bar{z})~.
\end{equation}
%The vertex operator
%$V^{\prime (0) \mathrm{DDF}}_{r}(Z_{r},\bar{Z}_{r})$
%is obtained from $V^{\prime \mathrm{DDF}}_{r} (Z_{r},\bar{Z}_{r})$
%by picture changing as 
%\begin{equation}
%V^{\prime (0) \mathrm{DDF}}_{r} (Z_{r},\bar{Z}_{r})
%= \left.\left(
%   \oint_{Z_{r}} \frac{dz}{2\pi i}
%    j_{\mathrm{B}} (z) \xi (Z_{r})
% \oint_{\bar{Z}_{r}} \frac{d\bar{z}}{2\pi i}
%   \tilde{\jmath}_{\mathrm{B}} (\bar{z}) \tilde{\xi} (\bar{Z}_{r})
%   e^{-\phi -\tilde{\phi}} V^{\prime \mathrm{DDF}}_{r}
%   (Z_{r},\bar{Z}_{r})\right)
%  \right|_{0}~,
%\end{equation}
%where $j_{\mathrm{B}} (z)$ is the BRST current
%and the symbol $~~|_{0}$ means the operation to pick up the terms 
%whose $\beta\gamma$-ghost number is $0$.
Total derivative terms with respect to the moduli parameters
$\mathcal{T}_{\mathcal{I}}$  arise
in rearranging $\mathcal{A}_{N}$ into the above form.
However, they vanish with $d$ largely negative
and the  external momenta $p_{r}^{\mu}$ appropriately chosen,
as explained above.
We define the amplitudes for such $d$ and analytically continue it
to $d=10$.
In the form of the amplitude given in eq.(\ref{eq:AN-BRSTinv}), 
the divergences corresponding to the limit $z_{I}\to z_{J}$
are no longer there for any value of $d$.
Therefore we can take the limit $d \rightarrow 10$
in this expression,
and it coincides with the result of the first quantized theory,
\begin{eqnarray}
\mathcal{A}_{N} 
  &\sim&
 \int \left[ dX d\psi d\tilde{\psi} d(\mbox{ghost}) \right]
        e^{-S_{d=10}}
\nonumber \\
&&  
\times
  \prod_{s=1,2}\left[
        c\tilde{c} e^{-\phi -\tilde{\phi}}
        V_{s}^{\mathrm{DDF}} (Z_{s},\bar{Z}_{s}) \right]
  c\tilde{c} V^{(0)\mathrm{DDF}}_{3} (Z_{3},\bar{Z}_{3})
  \prod_{r=4}^{N} 
       \int d^{2} Z_{r} 
       V^{(0)\mathrm{DDF}}_{r} (Z_{r},\bar{Z}_{r})~,~~~~
\end{eqnarray}
where
$S_{d=10}$ denotes the worldsheet action of the $d=10$ dimensional
NSR superstring with the ghosts,
which is obtained from $S$ in eq.(\ref{eq:actionS}) by setting
$d=10$.

%%%%%%%%%%%%%%%%%%%%%%%%%%%%%%%%%%%%%%%%%%%%
\section{Conclusions and Discussions}
  \label{sec:Conclusions-and-Discussions}

In this paper, we have formulated a dimensional regularization scheme to deal
with the divergences in the light-cone gauge 
closed string field theory for NSR superstrings. 
Starting from the action (\ref{eq:action}), 
we have obtained the tree level amplitudes with 
(NS,NS) external lines, 
which can be recast into a BRST invariant form using the 
superconformal field theory proposed in Ref.~\cite{Baba:2009fi}.
We have shown that the results coincide with those of the first quantized
formulation without introducing any contact term interactions. 

There are several things which remain to be done to show that our
scheme really works. 
One thing is to include the Ramond sector fields.
Another is to examine how to apply our dimensional regularization
to the multi-loop amplitudes. 
In dealing with the ultraviolet divergences in the loop amplitudes, 
the way to take the number of the Ramond sector ground states 
for $d\neq 10$ will be important. 
We may have to take something like the dimensional reduction scheme in 
supersymmetric field theory. 
We hope that we come back to these problems elsewhere. 

%%%%%%%%%%%%%%%%%%%%%%%%%%%%%%%%%%%%%%%%%%%%%%%%%%%
\section*{Acknowledgements}

N.I.\ and K.M.\ would like to thank the organizers of the 
workshop ``APCTP Focus Program on Current Trends in String 
Field Theory" at APCTP, Pohang, for the hospitality, 
where part of this work was done. 
This work was supported in part by 
Grant-in-Aid for Scientific Research~(C) (20540247)
and
Grant-in-Aid for Young Scientists~(B) (19740164) from
the Ministry of Education, Culture, Sports, Science and
Technology (MEXT).
%%%%%%%%%%%%%%%%%%%%%%%%%%%%%%%%%%%%%
\appendix

%%%%%%%%%%%%%%%%%%%%%%%%%%%%%%%%%%%%%%
\section{Action of Light-cone Gauge String Field Theory for $d\neq 10$}
\label{sec:action}

In this appendix,
we explain the details of the action (\ref{eq:action}) defined 
for $d\neq 10$. 

We represent the string field $\left|\Phi (t) \right\rangle $
by a Fock state for the non-zero modes and  
a wave function for the zero-modes $(t,\alpha,\vec{p})$, 
where $\alpha=2p^{+}$ is the string-length parameter
and $\vec{p}$ is the transverse $(d-2)$-momentum.
The integration measure $dr$ for the momentum zero-modes 
of the $r$-th string is defined as
\begin{equation}
dr =  \frac{\alpha_{r}d\alpha_{r}}{4\pi}
         \frac{d^{d-2}p_{r}}{\left(2\pi\right)^{d-2}}~.
\end{equation}
The string field $|\Phi (t)\rangle$ is taken to be GSO even
and satisfy the level matching condition:
\begin{equation}
\mathcal{P}_{\mathrm{GSO}} |\Phi (t) \rangle
 =|\Phi (t) \rangle~,
\qquad
\int^{2\pi}_{0} \frac{d\theta}{2\pi}
   e^{-i\theta 
      \left(L^{\mathrm{LC}}_{0}-\tilde{L}^{\mathrm{LC}}_{0}
      \right)}
|\Phi (t) \rangle = |\Phi (t)\rangle~,
\end{equation}
as well as the reality condition,
where $L_0^{\mathrm{LC}}$ denotes the zero-mode of the transverse
Virasoro generator.

In the action (\ref{eq:action}),
$g$ is the coupling constant.
$\left\langle R\left(1,2\right)\right|$ is the reflector
given by
\begin{eqnarray}
&&
\left\langle R\left(1,2\right)\right| 
=  \frac{1}{\alpha_{1}} \delta\left(1,2\right)
      {}_{12}\left\langle 0\right|e^{E\left(1,2\right)}~,
\qquad
{}_{12} \langle 0 |
 ={}_{2}\langle 0 |
  {}_{1}\langle 0|~,
\nonumber\\
&& \qquad E\left(1,2\right) 
 = 
 -\sum_{n=1}^{\infty} \frac{1}{n} 
     \left( \alpha_{n}^{i (1)} \alpha_{n}^{i (2)}
            +\tilde{\alpha}_{n}^{i (1)} \tilde{\alpha}_{n}^{i (2)}
     \right) 
 + i \sum_{r>0}
     \left( \psi_{r}^{i (1)} \psi_{r}^{i (2)}
            +\tilde{\psi}_{r}^{i (1)}\tilde{\psi}_{r}^{i(2)}\right)~,
\nonumber \\
&& \qquad \delta\left(1,2\right) 
  =  4\pi \delta \left(\alpha_{1}+\alpha_{2}\right)
        \left(2\pi\right)^{d-2} \delta\left(p_{1}+p_{2}\right)~.
\end{eqnarray}
$\left\langle V_{3}\left(1,2,3\right)\right|$
denotes the three-string interaction vertex defined as
\begin{eqnarray}
&& \left\langle V_{3}\left(1,2,3\right)\right| 
=  4 \pi \delta \Biggl( \sum_{r=1}^{3}\alpha_{r} \Biggr)
      \left(2\pi\right)^{d-2}
      \delta^{d-2} \Biggl( \sum_{r=1}^{3}p_{r} \Biggr)
      \left\langle V_{3}^{\mathrm{LPP}}\left(1,2,3\right)\right|
        P_{123} \, e^{-\Gamma^{[3]} \left(1,2,3\right)}~,
\nonumber\\
%&& \qquad
%\left\langle V_{3}^{\mathrm{LPP}}\left(1,2,3\right)\right| 
%  =  {}_{123}\left\langle 0\right|e^{E\left(1,2,3\right)}~,
%\qquad
%{}_{123} \langle 0|
% = {}_{3}\langle 0| {}_{2}\langle 0|{}_{1}\langle 0|~,
%\nonumber\\
&& \qquad
e^{-\Gamma^{[3]}\left(1,2,3\right)} 
   =  \mathrm{sgn}\left(\alpha_{1}\alpha_{2}\alpha_{3}\right)
      \left| \frac{e^{-2\hat{\tau}_{0}
                     \sum_{r} \frac{1}{\alpha_{r}}}}
                  {\alpha_{1}\alpha_{2}\alpha_{3}}
      \right|^{\frac{d-2}{16}}~,
\qquad
\hat{\tau}_{0} 
   =  \sum_{r=1}^{3} \alpha_{r} \ln \left| \alpha_{r} \right|~,
\end{eqnarray}
where 
$\left\langle V_{3}^{\mathrm{LPP}}\left(1,2,3\right)\right|$
is the LPP vertex~\cite{LeClair:1988sp}.
By the definition of the LPP vertex, for local operators 
$\mathcal{O}_{i}\left(\rho_{i},\bar{\rho}_{i}\right)$ on
the light-cone diagram, 
\begin{eqnarray}
 &  &
\int \frac{d^{d-2}p_{1}}{(2\pi)^{d-2}}
     \frac{d^{d-2}p_{2}}{(2\pi)^{d-2}}
     \frac{d^{d-2}p_{3}}{(2\pi)^{d-2}} \,
( 2\pi )^{d-2} \delta^{d-2} \Biggl( \sum_{r=1}^{3}p_{r} \Biggr)
\nonumber\\
&& \hspace{5em}
 \times
 \left\langle V_{3}^{\mathrm{LPP}}\left(1,2,3\right)\right|
   \mathcal{O}_{1}\left(\rho_{1},\bar{\rho}_{1}\right)
    \cdots
   \mathcal{O}_{n}\left(\rho_{n},\bar{\rho}_{n}\right)
\prod_{r=1}^{3}
\left( 
|0\rangle_{r} (2\pi)^{d-2} \delta^{d-2} (p_{r})
\right)
\nonumber\\
 &  & \qquad
  =\biggl\langle 
     \mathcal{O}_{1} \left( \rho \left(z_{1}\right),
                         \bar{\rho}\left(\bar{z}_{1}\right)\right)
    \cdots
    \mathcal{O}_{n} \left(\rho\left(z_{n}\right),
                         \bar{\rho}\left(\bar{z}_{n}\right)\right)
   \biggr\rangle~,
\end{eqnarray}
where $\rho (z)$ is the Mandelstam mapping~(\ref{eq:Mandelstam-N})
with $N=3$,
and $\langle \mathcal{O} \rangle$ is given in 
eq.(\ref{eq:expectation}).
The pre\-factor $P_{123}$ in the three-string vertex is defined
to satisfy
\begin{eqnarray}
&&\int \frac{d^{d-2}p_{1}}{(2\pi)^{d-2}}
     \frac{d^{d-2}p_{2}}{(2\pi)^{d-2}}
     \frac{d^{d-2}p_{3}}{(2\pi)^{d-2}} \,
  ( 2\pi )^{d-2} \delta^{d-2} \Biggl( \sum_{r=1}^{3}p_{r} \Biggr)
\nonumber\\
&& \hspace{5em}
  \times \left\langle V_{3}^{\mathrm{LPP}}\left(1,2,3\right)\right|
   P_{123} \, \mathcal{O}_{1}\left(\rho_{1},\bar{\rho}_{1}\right)
       \cdots \mathcal{O}_{n}\left(\rho_{n},\bar{\rho}_{n}\right)
  \prod_{r=1}^{3} 
  \left(
  |0\rangle_{r} (2\pi)^{d-2} \delta^{d-2} (p_{r})
  \right)
\nonumber \\
&& = \left(\partial^{2}\rho\left(z_{0}\right)
           \bar{\partial}^{2} \bar{\rho}\left(\bar{z}_{0}\right)
      \right)^{-\frac{3}{4}}
     \left\langle T_{F}^{\mathrm{LC}}\left(z_{0}\right)
     \tilde{T}_{F}^{\mathrm{LC}}\left(\bar{z}_{0}\right)
     \mathcal{O}_{1} \left(\rho\left(z_{1}\right),
                            \bar{\rho}\left(\bar{z}_{1} \right) 
                     \right)
     \cdots
     \mathcal{O}_{n}\left(\rho\left(z_{n}\right),
                           \bar{\rho}\left(\bar{z}_{n}\right)
                    \right)
     \right\rangle.~~~~~~
\label{eq:prefactor}
\end{eqnarray}
$z_{0}$ here denotes the $z$ coordinate of the interaction point
which satisfies $\partial\rho\left(z_{0}\right)=0$.

%%%%%%%%%%%%%%%
\section{Amplitudes\label{sec:Amplitudes} }

In this appendix, 
we calculate the tree level amplitudes perturbatively 
starting from the action (\ref{eq:action}). 
Here we calculate four-string amplitude explicitly as an example. 
It is straightforward to generalize the results to $N$-string case. 

\subsubsection*{Propagator and vertex}

It is convenient to introduce a basis
$\left\{ \left|n\right\rangle \right\} $
of the projected Fock space for the non-zero modes
which satisfies 
\begin{equation}
\left\langle n|n'\right\rangle   =  \delta_{n,n'}~,
\qquad
{}_{12}\left\langle 0 \right|
      e^{E\left(1,2\right)} \left|n\right\rangle _{1} 
=  {}_{2}\left\langle n\right|~,
\end{equation}
so that $\left|\Phi\right\rangle $ can be expanded as 
\begin{equation}
\left|\Phi (t)\right\rangle  
 =  \sum_{n}\phi_{n}\left(t,\alpha,\vec{p} \right)
        \left|n\right\rangle,
\end{equation}
and 
\begin{equation}
\left(L_{0}^{\mathrm{LC}} +\tilde{L}_{0}^{\mathrm{LC}}
      -\frac{d-2}{8}
\right)
\left| \Phi (t) \right\rangle  
 =  \sum_{n} \left( \vec{p}^{\; 2}+m_{n}^{2}\right)
      \phi_{n} \left(t, \alpha, \vec{p} \right)
      \left| n \right\rangle.
\end{equation}
$\phi_{n}$ corresponds to a particle in the spectrum of the string
and $m_{n}$ is the mass of the particle. 
The kinetic term of the action (\ref{eq:action}) can be rewritten as 
\begin{eqnarray}
&& \frac{1}{2}\int dt\int d1d2
   \left\langle R \left(1,2\right) | \Phi (t) \right\rangle _{1}
        \left( i\frac{\partial}{\partial t}
               -\frac{L_{0}^{\mathrm{LC}(2)}
                        +\tilde{L}_{0}^{\mathrm{LC}(2)}
                        -\frac{d-2}{8}}
                     {\alpha_{2}}
      \right)
     \left| \Phi (t) \right\rangle _{2}
\nonumber \\
 &  & \qquad
      =\int \frac{d^{d}p}{(2\pi)^{d}} \, 
       \sum_{n}\tilde{ \phi}_{n} (-p) 
           \left[ -\frac{1}{2} \left(p^{2}+m_{n}^{2}\right) \right]
         \tilde{\phi}_{n}(p)~,
\end{eqnarray}
where
\begin{equation}
\tilde{\phi}_{n} (p)
 \equiv \int dt \, e^{ip^{-}t} \phi_{n} (t,\alpha,\vec{p})~.
\end{equation}
Then we obtain the propagator
\begin{equation}
\wick{2}{<2{\tilde{ \phi}}_{n} \left(p\right) >2{\tilde{\phi}}_{n'}\left(p'\right)} 
=  \delta_{n,n'} \left(2\pi\right)^{d}
   \delta^{d} \left( p+p' \right)
   \frac{-i}{p^{2}+m_{n}^{2}}~.
\end{equation}
In terms of the string field 
$\Bigl| \tilde{\Phi} (p^{-}) \Bigr\rangle$,
defined as
\begin{equation}
\Bigl| \tilde{\Phi}\left(p^{-}\right) \Bigr\rangle  
\equiv  \int dt \, e^{ip^{-}t}
     \left|\Phi\left(t\right)\right\rangle
= \sum_{n} \tilde{\phi}_{n} (p) |n\rangle~,
\end{equation}
the propagator becomes
\begin{eqnarray}
\lefteqn{
\wick{2}{\Bigl| <2{\tilde{\Phi}} \left(p_{1}^{-}\right) 
         \Bigr\rangle_{1}
         \Bigl| >2{\tilde{\Phi}}_{2} \left(p_{2}^{-}\right) 
         \Bigr\rangle_{2}}
}\nonumber \\
&& =  \left(2\pi\right)^{d}
         \delta^{d} \left( p_{1}+p_{2} \right)
         \sum_{n}
         \frac{-i}{p_{1}^{2}+m_{n}^{2}}
      \left|n\right\rangle _{1}\left|n\right\rangle _{2}
\nonumber\\
&& = 
   -i\left(2\pi\right)^{d} \delta^{d} \left(p_{1}+p_{2}\right)
   \frac{1}{\left|\alpha_{1}\right|}  \sum_{n} 
   \int_{0}^{\infty} dT \, 
          e^{-\frac{T}{\left|\alpha_{1}\right|}
                \left(p_{1}^{2}+m_{n}^{2}\right)}
          \left| n \right\rangle _{1}
          \left|n\right\rangle _{2}
\nonumber \\
% && =  -i\left(2\pi\right)^{d}
%       \delta^{d} \left(p_{1}+p_{2}\right)
%       \frac{1}{\left|\alpha_{1}\right|}
%       \int_{0}^{\infty}dT \,
%          e^{-\frac{T}{\left|\alpha_{1}\right|}
%               \left(-\alpha_{1}p_{1}^{-}+L_{0}^{\mathrm{LC} (1)}
%                      +\tilde{L}_{0}^{\mathrm{LC} (1)}
%                      -\frac{c}{12}\right)}
%\nonumber\\
% && 
% \hphantom{
%       =
%       -i\left(2\pi\right)^{d}\delta^{d}\left(p_{1}+p_{2}\right)
%       \frac{1}{\left|\alpha_{1}\right|}
%           }
%       \times \int_{0}^{2\pi} \frac{d\theta}{2\pi} \,
%         e^{-i\theta\left(L_{0}^{\mathrm{LC}(1)}
%                          -\tilde{L}_{0}^{\mathrm{LC} (1)}\right)}
%        \mathcal{P}_{\mathrm{GSO}}^{(1)}
%        \mathcal{P}_{\mathrm{GSO}}^{(2)}
%        \, e^{E\left(1,2\right)}  \left| 0 \right\rangle _{12}
%\nonumber \\
 && =  \frac{1}{\alpha_{1}}\left(2\pi\right)^{d}
       \delta^{d} \left(p_{1}+p_{2}\right)
       \int \frac{d^{2} \mathcal{T}}{4\pi} \,
          e^{-\frac{\mathcal{T}}{\left|\alpha_{1}\right|}
                \left(L_{0}^{\mathrm{LC} (1)}-\frac{d-2}{16}\right)
             -\frac{\bar{\mathcal{T}}}{\left|\alpha_{1}\right|}
                \left(\tilde{L}_{0}^{\mathrm{LC} (1)}
                    -\frac{d-2}{16}\right)}
\nonumber\\
 && 
  \hphantom{
    =
    \frac{-i}{\alpha_{1}}\left(2\pi\right)^{d}
       \delta^{d}\left(p_{1}+p_{2}\right)
       \int\frac{d^{2}\mathcal{T}}{4\pi}
    }
\, \times 
   e^{\frac{\alpha_{1}}{\left|\alpha_{1}\right|} p_{1}^{-}T}
   \frac{1}{\alpha_{1}}
   \mathcal{P}_{\mathrm{GSO}}^{(1)}
   \mathcal{P}_{\mathrm{GSO}}^{(2)} \,
   e^{E^{\dagger} \left(1,2\right)} \left| 0 \right\rangle _{12}~,
\end{eqnarray}
where 
\begin{eqnarray}
\mathcal{T} &\equiv&  T+i\left|\alpha_{1}\right|\theta~,
\nonumber\\
\int\frac{d^{2}\mathcal{T}}{4\pi} 
 &\equiv& 
 -i\left| \alpha_{1} \right|
 \int_{0}^{\infty}dT \int_{0}^{2\pi}
 \frac{d\theta}{2\pi}~.
\end{eqnarray}
The Schwinger parameter $\mathcal{T}$ will become
a complex moduli parameter of the amplitudes.
Another useful form of the propagator is 
\begin{equation}
\wick{2}{<2{\tilde{\phi}}_{n} \left(p\right) 
\Bigl| >2{\tilde{\Phi}} \left(p'\right) \Bigr\rangle}
=  \left(2\pi\right)^{d} \delta^{d} \left(p+p'\right)
   \frac{-i}{p^{2}+m_{n}^{2}} \left| n \right\rangle~.
\label{eq:Schwinger}
\end{equation}

In terms of $|\tilde{\Phi}\left(p^{-}\right) \rangle $, 
the three-string interaction term can be written as 
\begin{eqnarray}
\lefteqn{
\int dt\int d1d2d3\left\langle V_{3}\left(1,2,3\right)
   |\Phi (t) \right\rangle _{1}
   \left| \Phi (t) \right\rangle _{2}
   \left| \Phi (t) \right\rangle _{3}
}\nonumber \\
 &&= \int \prod_{r=1}^{3}
           \left(  \frac{d^{d}p_{r}}{\left(2\pi\right)^{d}}
                  \alpha_{r}
           \right)
    \left(2\pi\right)^{d} \delta^{d} \Biggl(\sum_{r=1}^{3}p_{r}\Biggr)
    e^{-\Gamma^{\left[3\right]}\left(1,2,3\right)}
\nonumber \\
 &  & \hphantom{=\prod \int}
      \qquad \qquad \,
      \times 
      \left\langle V_{3}^{\mathrm{LPP}} \left(1,2,3\right) \right|
         P_{123} 
         \Bigl|\tilde{\Phi}\left(p_{1}^{-}\right)\Bigr\rangle _{1}
         \Bigl|\tilde{\Phi}\left(p_{2}^{-}\right)\Bigr\rangle _{2}
         \Bigl|\tilde{\Phi}\left(p_{3}^{-}\right)\Bigr\rangle _{3}~.
\label{eq:3vertex}
\end{eqnarray}

\subsubsection*{Four-string amplitudes}

The four-string amplitudes $\mathcal{A}_{4}$
can be calculated from the correlation functions 
of the string field theory,
\begin{equation}
\left\langle \! \left\langle
  \tilde{\phi}_{n_{1}} (p_{1}) \, \tilde{\phi}_{n_{2}} (p_{2}) \,
  \tilde{\phi}_{n_{3}} (p_{3}) \, \tilde{\phi}_{n_{4}} (p_{4}) 
\right\rangle \! \right\rangle~,
\end{equation}
which can be calculated perturbatively 
by using the three-string vertex in eq.(\ref{eq:3vertex}). 
Here $\langle\! \langle \cdots \rangle\!\rangle$ denotes 
the expectation value 
in the string field theory. 
The tree level contribution becomes 
\begin{eqnarray}
&& \left(4ig\right)^{2}\left(2\pi\right)^{d}
    \delta^{d} \Biggl( \sum_{r=1}^{4}p_{r} \Biggr)
    \prod_{r=1}^{4} \left(\frac{-i}{p_{r}^{2}+m_{n_{r}}^{2}}
                          \alpha_{r}\right)
\nonumber\\
&& \quad \times
   \left[ -\int\frac{d^{2}\mathcal{T}}{4\pi}
         \left\langle V_{3}^{\mathrm{LPP}}\left(1,2,5\right)\right|
         \left\langle V_{3}^{\mathrm{LPP}}\left(3,4,6\right)\right|
         P_{125}P_{346}
    \right.
\nonumber\\
&& \hphantom{
       \quad\times\quad-\int\frac{d^{2}\mathcal{T}}{4\pi}
       }
\times e^{-\frac{\mathcal{T}}{\left| \alpha_{5} \right|}
           \left( L_{0}^{\mathrm{LC} (5)}-\frac{d-2}{16}\right)
          -\frac{\bar{\mathcal{T}}}{\left|\alpha_{5}\right|}
           \left(\tilde{L}_{0}^{\mathrm{LC} (5)}-\frac{d-2}{16}\right)}
       \,
       e^{\frac{\alpha_{5}}{\left|\alpha_{5}\right|}p_{5}^{-}T}
\nonumber \\
&& \hphantom{
        \quad\times\quad-\int\frac{d^{2}\mathcal{T}}{4\pi}
        }
\times \mathcal{P}_{\mathrm{GSO}}^{(5)}
       \mathcal{P}_{\mathrm{GSO}}^{(6)} \, 
       e^{E^{\dagger}\left(5,6\right)}
       \left| 0 \right\rangle _{56}
       \left| n_{1} \right\rangle _{1} \left| n_{2} \right\rangle _{2}
       \left| n_{3} \right\rangle _{3} \left| n_{4} \right\rangle _{4}
       \, e^{-\Gamma^{[3]} \left(1,2,5\right)}
          e^{-\Gamma^{[3]} \left(3,4,6\right)}
\nonumber\\
 && \hphantom{\quad\times\quad}
    \left.
     {}+\mbox{other channels}
     \vphantom{\int\frac{d^{2}\mathcal{T}}{4\pi}}
    \right].
\end{eqnarray}
%where $P_{125}$ and $P_{346}$ denote the prefactors for 
%the three-string vertices defined in eq.(\ref{eq:prefactor}).

The amplitudes $\mathcal{A}_{4}$ can be obtained from 
the correlation functions
by amputating the external legs 
and putting $p_{r}$ on the mass shell:
\begin{eqnarray}
0 & = & p_{r}^{2}+m_{n_{r}}^{2}\nonumber \\
 & = & {}-2p_{r}^{+}p_{r}^{-} + \vec{p}_{r}^{\;2}
       +m_{n_{r}}^{2}~.
\label{eq:on-shell}
\end{eqnarray}
At the tree level it can therefore be written as
\begin{equation}
\mathcal{A}_{4}  
=  \left(4ig\right)^{2}
   \left[ \int\frac{d^{2}\mathcal{T}}{4\pi}\,
          F_{4} \left(\mathcal{T},\bar{\mathcal{T}}\right)
          +\mbox{other channels}\right],
\label{eq:amplitude4}
\end{equation}
where
\begin{eqnarray}
F_{4} \left(\mathcal{T},\bar{\mathcal{T}}\right)
& \equiv &
   -\left(2\pi\right)^{d}
    \delta^{d} \Biggl( \sum_{r=1}^{4}p_{r} \Biggr)
    \left(\prod_{r=1}^{4}\alpha_{r}\right)
    e^{-\Gamma^{[3]} \left(1,2,5\right)}
    e^{-\Gamma^{[3]} \left(3,4,6\right)}
\nonumber \\
&& \quad \times
       \left\langle V_{3}^{\mathrm{LPP}}\left(1,2,5\right)\right|
       \left\langle V_{3}^{\mathrm{LPP}}\left(3,4,6\right)\right|
    P_{125}P_{346}
\nonumber \\
&& \quad \times 
   e^{-\frac{\mathcal{T}}{\left|\alpha_{5}\right|}
        \left( L_{0}^{\mathrm{LC} (5)}-\frac{d-2}{16} \right)
      -\frac{\bar{\mathcal{T}}}{\left|\alpha_{5}\right|}
         \left(\tilde{L}_{0}^{\mathrm{LC} (5)}
                -\frac{d-2}{16}\right)} \,
   e^{\frac{\alpha_{5}}{\left|\alpha_{5}\right|}p_{5}^{-}T}
\nonumber \\
 &  & \quad \times 
    e^{E^{\dagger} \left(5,6\right)} \left|0\right\rangle _{56}
    \left| n_{1} \right\rangle _{1} \left| n_{2} \right\rangle _{2}
    \left| n_{3} \right\rangle _{3} \left| n_{4} \right\rangle _{4}~.
\end{eqnarray}
The integrand $F_{4}\left(\mathcal{T},\bar{\mathcal{T}}\right)$
corresponds to a light-cone diagram for the four-string amplitude.
The light-cone diagram can be mapped to the complex $z$-plane 
by the Mandelstam mapping $\rho (z)$
in eq.(\ref{eq:Mandelstam-N}) with $N=4$.
For later use, 
for each of the regions $z\sim Z_{r}$ $(r=1,\ldots,4)$
to which the external lines are mapped by
the Mandelstam mapping $\rho (z)$,
we introduce the local coordinate $w_{r}$ defined as
\begin{equation}
w_{r} 
\equiv 
 \exp \left[ \frac{1}{\alpha_{r}}
             \left(\rho-\tau_{0}^{\left(r\right)}-i\beta_{r}\right)
      \right].
\label{eq:def-wr}
\end{equation} 
Here $\tau_{0}^{(r)}+i\beta_{r}$
are given in eq.(\ref{eq:Nbar-rr-00}). 
%%The two interaction points are mapped to $z_{I}$
%%$\left(I=1,2\right)$
%%which satisfy $\partial\rho \left(z_{I}\right)=0$. 
The Schwinger parameter $\mathcal{T}$ is expressed 
as the difference between the $\rho\left(z_{I}\right)$'s.
It is easy to see
\begin{equation}
\frac{\alpha_{5}}{\left|\alpha_{5}\right|}p_{5}^{-}T
 =  -\sum_{r=1}^{4}p_{r}^{-}\tau_{0}^{\left(r\right)}~.
\end{equation}
Via the Mandelstam mapping, 
$F\left(\mathcal{T},\bar{\mathcal{T}}\right)$
can be expressed in terms of the correlation functions
of the worldsheet theory
on the complex  $z$-plane as 
\begin{eqnarray}
F_{4}\left(\mathcal{T},\bar{\mathcal{T}}\right)
 &=& \left(2\pi\right)^{2}
     \delta \Biggl( \sum_{r=1}^{4} p_{r}^{+} \Biggr)
     \delta \Biggl( \sum_{r=1}^{4} p_{r}^{-} \Biggr)
     e^{-\Gamma^{\left[4\right]}\left(1,2,3,4\right)}
\nonumber \\
 & & \quad \times
     \left\langle \prod_{I=1,2}
         \left[ \left(\partial^{2}\rho\left(z_{I}\right)
                      \bar{\partial}^{2} \bar{\rho}
                             \left(\bar{z}_{I} \right)
                \right)^{-\frac{3}{4}}
                T_{F}^{\mathrm{LC}} \left(z_{I}\right)
                \tilde{T}_{F}^{\mathrm{LC}}\left(\bar{z}_{I}\right)
         \right]
     \prod_{r=1}^{4}V_{r}^{\mathrm{LC}}
\right\rangle,
\label{eq:FTTbar}
\end{eqnarray}
where 
\begin{eqnarray}
&&e^{-\Gamma^{\left[4\right]}\left(1,2,3,4\right)} 
= -e^{-\Gamma^{\left[3\right]}\left(1,2,5\right)}
     e^{-\Gamma^{\left[3\right]}\left(3,4,6\right)}
   \left\langle V_{3}^{\mathrm{LPP}}\left(1,2,5\right)\right|
   \left\langle V_{3}^{\mathrm{LPP}}\left(3,4,6\right)\right|
\nonumber\\
&& \hphantom{ 
            e^{-\Gamma^{\left[4\right]}\left(1,2,3,4\right)} 
            }
   \quad \times 
e^{-\frac{\mathcal{T}}{\left|\alpha_{5}\right|}
      \left(L_{0}^{\mathrm{LC}(5)}-\frac{d-2}{16}\right)
   -\frac{\bar{\mathcal{T}}}{\left|\alpha_{5}\right|}
     \left(\tilde{L}_{0}^{\mathrm{LC} (5)}-\frac{d-2}{16}\right)}
e^{E^{\dagger} \left(5,6\right)}
\left| 0 \right\rangle_{56}
\left| 0 \right \rangle_{1} \left| 0 \right\rangle_{2} 
\left| 0 \right \rangle_{3} \left| 0 \right\rangle_{4},~~~~~~
\end{eqnarray}
and the vertex operator $V^{\mathrm{LC}}_{r}$ is defined as
\begin{eqnarray}
V_{r}^{\mathrm{LC}}
 &=& \alpha_{r}
%%\epsilon^{(r)}_{\{ij\}}
     \frac{i\partial^{n_{1}}X^{i_{1}}\left(w_{r}\right)}
          {\left(n_{1}-1\right)!}
     \cdots
     \frac{i\bar{\partial}^{\tilde{n}_{1}}X^{\tilde{\imath}_{1}}
             \left(\bar{w}_{r}\right)}
          {\left(\tilde{n}_{1}-1\right)!}
     \cdots
     \frac{\partial^{s_{1}-\frac{1}{2}} \psi^{j_{1}}
             \left(w_{r}\right)}
          {\left(s_{1}-\frac{1}{2}\right)!}
     \cdots
     \frac{\bar{\partial}^{\tilde{s}_{1}-\frac{1}{2}}
            \tilde{\psi}^{\tilde{\jmath}_{1}}
            \left(\bar{w}_{r}\right)}
          {\left(\tilde{s}_{1}-\frac{1}{2}\right)!}
     \cdots
\nonumber \\
 &  & \hphantom{\alpha_{r}}
      \left.\times e^{ip_{r}^{i}X^{i}}
      \left(w_{r},\bar{w}_{r}\right)\right|_{w_{r}=\bar{w}_{r}=0}
      \; e^{-p_{r}^{-}\tau_{0}^{\left(r\right)}}~,
\label{eq:VLC}
\end{eqnarray}
corresponding to the state whose non-zero mode part is
$|n_{r}\rangle_{r}$, namely
\begin{equation}
|n_{r} \rangle_{r}
=
%%\epsilon^{(r)}_{\{ ij \}} 
   \alpha_{-n_{1}}^{i_{1} (r)}
        \cdots
     \tilde{\alpha}_{-\tilde{n}_{1}}^{\tilde{\imath}_{1} (r)}
       \cdots
     \psi_{-s_{1}}^{j_{1} (r)}
       \cdots
     \tilde{\psi}_{-\tilde{s}_{1}}^{\tilde{\jmath}_{1} (r)}
       \cdots
     \left| 0 \right\rangle _{r}~,
\label{eq:lcgaugestate}
\end{equation}
up to a normalization constant.
%%Here $\epsilon^{(r)}_{\{ij\}}$ is a shorthand notation
%%of the polarization tensor
%%$\epsilon^{(r)}_{i_{1}\cdots \tilde{\imath}_{1}\cdots
%%                 j_{1}\cdots \tilde{\jmath}_{1} \cdots}
%%$
%%of the vertex operator.
$e^{-\Gamma^{\left[4\right]}\left(1,2,3,4\right)}$ is the partition
function for the four-string light-cone diagram
and should behave as 
\begin{equation}
e^{-\Gamma^{\left[4\right]}\left(1,2,3,4\right)}
\sim  -e^{-\Gamma^{\left[3\right]}\left(1,2,5\right)}
       e^{-\Gamma^{\left[3\right]}\left(3,4,6\right)}
       e^{\frac{d-2}{8}\frac{T}{\left|\alpha_{5}\right|}}
\label{eq:factorization}
\end{equation}
for $T=\mathop{\mathrm{Re}}\mathcal{T}\to\infty$. From these 
properties, one can show that
\begin{equation}
e^{-\Gamma^{\left[4\right]}\left(1,2,3,4\right)}
= \mathrm{sgn}\left(\alpha_{1}\alpha_{2}\alpha_{3}\alpha_{4}\right)
  e^{-\frac{d-2}{16} 
      \Gamma \left[ 
               \ln \left(\partial\rho \bar{\partial}\bar{\rho}
                   \right)
              \right]}~,
\label{eq:Gamma4}
\end{equation}
where 
$\Gamma \left[ \ln \left( \partial \rho \bar{\partial}\bar{\rho}
                   \right) \right]$
is given in eq.(\ref{eq:GammaN}) with $N=4$.

One of the most important properties of 
$F_{4} \left(\mathcal{T},\bar{\mathcal{T}}\right)$
is that the integrands in the other channels are obtained 
by analytically continuing $\mathcal{T}$. 
In order to show this property, we should prove that
$F_{4}(\mathcal{T},\bar{\mathcal{T}})$
in eq.(\ref{eq:FTTbar}) is independent of
$z_{I^{(r)}}$, because
the identification of $z_{I^{(r)}}$ depends on
the channel as explained below eq.(\ref{eq:Nbar-rr-00}).
%The key to the proof is that
%$F_{4}(\mathcal{T},\bar{\mathcal{T}})$
%in eq.(\ref{eq:FTTbar}) is independent of
%$\tau_{0}^{(r)}$ $(r=1,\ldots,4)$, which are discontinuous
%with respect to $\mathcal{T}$.
%This can be readily shown by using the facts that
%the $\tau_{0}^{(r)}$ dependent part of
%$e^{-\Gamma^{[4]}(1,2,3,4)}$ in eq.(\ref{eq:Gamma4})
%is $e^{-\frac{d-2}{8} \mathop{\mathrm{Re}} \bar{N}^{rr}_{00}}$,
%the image $V_{r}^{\mathrm{LC}} (Z_{r},\bar{Z}_{r})$
%of $V_{r}^{\mathrm{LC}}$ in eq.(\ref{eq:VLC}) 
%to the $z$-plane via the Mandelstam mapping is given by
Using the fact that $V_{r}^{\mathrm{LC}}$ in eq.(\ref{eq:VLC}) 
can be rewritten as 
\begin{equation}
V_{r}^{\mathrm{LC}}
 = \left| \frac{\partial w_{r}}{\partial z} (Z_{r})
   \right|^{-\left( \vec{p}_{r}^{2}+2N_{r} \right)}
  V^{\mathrm{LC}}_{r} (Z_{r},\bar{Z}_{r})
 = e^{\left( \frac{d-2}{8} + 2p^{+}_{r}p^{-}_{r} \right)
     \mathop{\mathrm{Re}} \bar{N}^{rr}_{00}}
 V_{r}^{\mathrm{LC}}(Z_{r},\bar{Z}_{r})~,
\end{equation}
where $V_{r}^{\mathrm{LC}}(Z_{r},\bar{Z}_{r})$ 
is the primary field corresponding to $V_{r}^{\mathrm{LC}}$ 
on the $z$-plane, 
it is easy to see that 
$F_{4}\left(\mathcal{T},\bar{\mathcal{T}}\right)$ 
%%in eq.(\ref{eq:FTTbar}) 
is independent of $z_{I^{(r)}}$ 
if all the external lines are on shell, 
and thus
depends only on the shape of the diagram. 
%and $V_{r}^{\mathrm{LC}} (Z_{r},\bar{Z}_{r})$ is
%proportional to $e^{-p_{r}^{-} \tau^{(r)}_{0}}$.
%Here $\bar{N}^{rr}_{00}$ denotes the Neumann coefficient
%(\ref{eq:Nbar-rr-00}), and
%we have used eqs.(\ref{eq:on-shell-2}) and (\ref{eq:def-wr}).
%Because of  Wick's theorem, the correlation function
%defined in eq.(\ref{eq:expectation}) can be expressed in terms of
%the propagators for $X^{i}$, $\psi^{i}$ and $\tilde{\psi}^{i}$ on the 
%worldsheet. 
%These propagators are essentially determined by
%the shape of the string diagram. 
%$F_{4}(\mathcal{T},\bar{\mathcal{T}})$ therefore
%depends on the shape of the diagram continuously.
Since $\Gamma^{[4]}$ given in eq.(\ref{eq:Gamma4}) satisfies 
the factorization property in eq.(\ref{eq:factorization}) 
for any channels, 
one can conclude that the expression (\ref{eq:FTTbar}) is valid 
for any channels, and thus the integrands in various channels are 
related by analytic continuation.
Therefore eq.(\ref{eq:amplitude4}) can be rewritten as 
\begin{equation}
\mathcal{A}_{4}
  =  \left(4ig\right)^{2}
     \int \frac{d^{2}\mathcal{T}}{4\pi} \,
     F_{4}\left(\mathcal{T},\bar{\mathcal{T}}\right),
\label{eq:A4}
\end{equation}
where now the integration region is taken to cover
the whole moduli space.
Hence, with the action (\ref{eq:action}), 
the amplitude can be expressed as an integral 
over the whole moduli space,
even in $d\neq10$ dimensional spacetime.
What is essential is the choice of 
$e^{-\Gamma^{[3]}\left(1,2,3\right)}$. 

It is straightforward to generalize the above procedure 
to show that $N$-string tree level amplitudes can be expressed 
as eqs.(\ref{eq:AN}) and (\ref{eq:FN0}).

%%%%%%%%%%%%%%%%%%%%%%%%%%%%%%%%%%%%%%%%%%%%%%
\section{Correlation Functions of $\psi^{-}$}
\label{sec:psiminus}

In this appendix, extracting the $X^{\pm}$ CFT part of the path integral
(\ref{eq:FN4}), we will prove that the terms of the form 
\begin{eqnarray}
 &  & \int[dX^{\pm}d\psi^{\pm}d\tilde{\psi}^{\pm}]
  e^{-S_{\pm}}
  \prod_{i=1}^{n}R\left(z_{I_{i}}\right)
  \prod_{j=1}^{\tilde{n}}\tilde{R}\left(\bar{z}_{I_{j}}\right)
\nonumber \\
 &  & \quad\times\prod_{r=1}^{N}
  \left[ e^{-ip_{r}^{+}X^{-}}(Z_{r},\bar{Z}_{r})
         \oint_{z_{I^{(r)}}}\frac{d\mathbf{z}}{2\pi i}
           D\Phi(\mathbf{z})
         \oint_{\bar{z}_{I^{(r)}}}\frac{d\bar{\mathbf{z}}}{2\pi i}
           \bar{D}\Phi(\bar{\mathbf{z}})
         e^{\frac{d-10}{16}\frac{i}{p_{r}^{+}}\mathcal{X}^{+}}
             (\mathbf{z},\bar{\mathbf{z}})
   \right]\ ,
\label{eq:appidentity}
\end{eqnarray}
vanish for any $n$ and $\tilde{n}$ 
with $1 \leq n,\tilde{n} \leq N-2$,
and for an arbitrary set of $n$ ($\tilde{n}$) distinct
interaction points
$z_{I_{i}}$ $\left(i=1,\ldots, n\right)$
($\bar{z}_{I_{j}}$ $\left(j=1,\ldots,\tilde{n}\right)$)
chosen out of $N-2$ $z_{I}$'s ($\bar{z}_{I}$'s). 
Using this fact, one can easily show that 
$R\left(z_{I}\right)$ and $\tilde{R}\left(\bar{z}_{I}\right)$
do not contribute to the correlation function (\ref{eq:Rz}) 
and the right hand side of eq.(\ref{eq:FN3}) coincides with that of 
eq.(\ref{eq:FN2}).

Since $X^{-}$ appears only in $S_{\pm}$ and 
$e^{-ip_{r}^{+}X^{-}}(Z_{r},\bar{Z}_{r})$,
one can see that $X^{+}$ in eq.(\ref{eq:appidentity}) can be replaced
by its expectation value $-\frac{i}{2}\left(\rho+\bar{\rho}\right)$.
The insertions at $z_{I^{(r)}}$ can be transformed as
\begin{eqnarray}
 &  & \oint_{z_{I^{(r)}}}\frac{d\mathbf{z}}{2\pi i}
        D\Phi(\mathbf{z})
       \oint_{\bar{z}_{I^{(r)}}}\frac{d\bar{\mathbf{z}}}{2\pi i}
        \bar{D}\Phi(\bar{\mathbf{z}})
       e^{\frac{d-10}{16}\frac{i}{p_{r}^{+}}\mathcal{X}^{+}}
         (\mathbf{z},\bar{\mathbf{z}})
\nonumber \\
 &  & \quad
 \sim \left(1+K\right) \left(1+\tilde{K}\right)
      e^{\frac{d-10}{16}\frac{1}{2p_{r}^{+}}
          \left(\rho+\bar{\rho}\right)}
      (z_{I^{(r)}},\bar{z}_{I^{(r)}})\ ,
\end{eqnarray}
where $K$ $(\tilde{K})$ consists of terms which involve derivatives
of $\psi^{+}$ $(\tilde{\psi}^{+})$. Therefore what we should show
is 
\begin{eqnarray}
 &  & \int[dX^{\pm}d\psi^{\pm}d\tilde{\psi}^{\pm}]e^{-S_{\pm}}
   \prod_{i=1}^{n}\oint_{z_{I_{i}}}\frac{dw_{i}}{2\pi i}
         \frac{\partial\rho\psi^{-}(w_{i})}{w_{i}-z_{I_{i}}}
   \prod_{j=1}^{\tilde{n}}
          \oint_{\bar{z}_{I_{j}}} \frac{d\bar{u}_{j}}{2\pi i}
          \frac{\bar{\partial}\bar{\rho}\tilde{\psi}^{-}(\bar{u}_{j})}
               {\bar{u}_{j}-\bar{z}_{I_{j}}}
\nonumber \\
 &  & \qquad\qquad\times
       \prod_{r=1}^{N}\left[
          e^{-ip_{r}^{+}X^{-}}(Z_{r},\bar{Z}_{r})
          \left(1+K(z_{I^{(r)}})\right)
          \left(1+\tilde{K}(\bar{z}_{I^{(r)}})\right)
       \right]
\nonumber \\
 &  & =0\ .
\label{eq:appidentity2}
\end{eqnarray}

Since $\partial\rho\left(w_{i}\right)=0$ at $w_{i}=z_{I_{i}}$, the
contour integral with respect to $w_{i}$ is nonvanishing only when
$\psi^{-}\left(w_{i}\right)$ has a singularity at $w_{i}=z_{I_{i}}$.
Here let us direct our attention to the variable $w_1$ 
and examine the singularities at $w_{1}=z_{I_{1}}$,
using the properties of the correlation functions in the $X^{\pm}$
CFT~\cite{Baba:2009fi}. 
Some of such singularities can come from the contraction
of $\psi^{-}\left(w_{1}\right)$ with a derivative of $\psi^{+}$
contained in $K(z_{I^{(r)}})$ 
such that $z_{I^{(r)}}=z_{I_{1}}$.
However, since $K$ consists of even number of $\psi^{+}$, such a
term necessarily involves another contraction of 
$\partial^{k-1}\psi^{+}\left(z_{I_{1}}\right)$ $\left(k\geq1\right)$
and $\psi^{-}\left(w_{i}\right)$ $\left(i\ne1\right)$, 
which is proportional to $\left(w_{i}-z_{I_{1}}\right)^{-k}$. 
Then the contour integral
of it over $w_{i}$ around $z_{I_{i}}(\neq z_{I_1})$ vanishes. 
Therefore such contractions do not contribute to the path integral 
in eq.(\ref{eq:appidentity2}).
The same arguments hold for the anti-holomorphic part.

Therefore we can ignore $K$ and $\tilde{K}$ 
in eq.(\ref{eq:appidentity2})
and what we should show becomes 
\begin{equation}
\int[dX^{\pm}d\psi^{\pm}d\tilde{\psi}^{\pm}]
  e^{-S_{\pm}}
  \prod_{r=1}^{N}e^{-ip_{r}^{+}X^{-}}(Z_{r},\bar{Z}_{r})
  \prod_{i=1}^{n}\oint_{z_{I_{i}}}\frac{dw_{i}}{2\pi i}
         \frac{\partial\rho\psi^{-}(w_{i})}{w_{i}-z_{I_{i}}}
  \prod_{j=1}^{\tilde{n}}\oint_{\bar{z}_{I_{j}}}
          \frac{d\bar{u}_{j}}{2\pi i}
         \frac{\bar{\partial}\bar{\rho}\tilde{\psi}^{-}(\bar{u}_{j})}
              {\bar{u}_{j}-\bar{z}_{I_{j}}}=0\ .
\label{eq:appidentity3}
\end{equation}
Now the problem is to examine the singularity of the correlation
function 
\begin{equation}
\int[dX^{\pm}d\psi^{\pm}d\tilde{\psi}^{\pm}]
  e^{-S_{\pm}}
  \prod_{r=1}^{N} e^{-ip_{r}^{+}X^{-}}(Z_{r},\bar{Z}_{r})
  \prod_{i=1}^{n}\psi^{-}(w_{i})\ ,
\label{eq:correlation-psi2}
\end{equation}
as a function of $w_{i}$. One can see that the contour integrals
over $w_{i}$ $\left(i=1,\cdots,n\right)$ in eq.(\ref{eq:appidentity3})
yield a nonvanishing result, 
only if the correlation function (\ref{eq:correlation-psi2})
behaves as 
\begin{eqnarray}
 &  & \int[dX^{\pm}d\psi^{\pm}d\tilde{\psi}^{\pm}]
       e^{-S_{\pm}}
       \psi^{-}(w_{1}) \cdots \psi^{-}(w_{n})
       \prod_{r=1}^{N} e^{-ip_{r}^{+}X^{-}}(Z_{r},\bar{Z}_{r})
\nonumber \\
 &  & \qquad\sim
   \left(w_{1}-z_{I_{1}}\right)^{-m_{1}}
   \left(w_{2}-z_{I_{2}}\right)^{-m_{2}} \cdots
   \left(w_{n}-z_{I_{n}}\right)^{-m_{n}}~,
\label{eq:correlation-psi}
\end{eqnarray}
for 
$\left(w_{1},w_{2},\cdots,w_{n}\right)
   \sim\left(z_{I_{1}},z_{I_{2}},\cdots,z_{I_{n}}\right)$,
where $m_{i}$ $(i=1,\cdots,n)$ are positive integers. 
Here $z_{I_{i}}$ should be all distinct in order to contribute 
to the correlation function (\ref{eq:appidentity}). 
In the following, we would like to show that
the correlation functions of $\psi^{-}$ cannot have 
the singularities of the form (\ref{eq:correlation-psi})
satisfying such conditions.

In the following analysis, it is convenient to 
introduce \cite{Baba:2009fi}
\begin{eqnarray}
\left\langle F[\mathcal{X}^{+},\mathcal{X}^{-}]
\right\rangle _{\rho}
  \equiv
  \frac{\int[d\mathcal{X}^{\pm}]e^{-S_{\pm}}
          F[\mathcal{X}^{+},\mathcal{X}^{-}]
          \prod_{r=1}^{N}e^{-ip_{r}^{+}\mathcal{X}^{-}}
                           (\mathbf{Z}_{r},\bar{\mathbf{Z}}_{r})}
       {\int[d\mathcal{X}^{\pm}]e^{-S_{\pm}}
           \prod_{r=1}^{N}e^{-ip_{r}^{+}\mathcal{X}^{-}}
                           (\mathbf{Z}_{r},\bar{\mathbf{Z}}_{r})}~.
\end{eqnarray}
Here 
$\mathbf{Z}_{r}=(Z_{r},\Theta_{r})$ and the subscript $\rho$ on
the left hand side stands for the super Mandelstam mapping 
$\rho(\mathbf{z})
   =\sum_{r=1}^{N}\alpha_{r}\ln(\mathbf{z}-\mathbf{Z}_{r})$.
Using this notation, the correlation function 
in eq.(\ref{eq:correlation-psi})
is proportional to 
\begin{equation}
\left.\left\langle 
   D\mathcal{X}^{-}(\mathbf{w}_{1})
        \cdots D\mathcal{X}^{-}(\mathbf{w}_{n})
       \right\rangle _{\rho}
\right|_{\theta_{i}=\Theta_{r}=0}~,
\label{eq:DXminus}
\end{equation}
where $\mathbf{w}_{i}=(w_{i},\theta_{i})$. 
As explained in Ref.~\cite{Baba:2009fi},
one can evaluate eq.(\ref{eq:DXminus}) starting from the one point
function $\langle D\mathcal{X}^{-}(\mathbf{w}_{1})\rangle_{\rho}$.
In order to do so, 
we introduce the super Mandelstam mapping $\rho_{m}$ defined as 
\begin{equation}
\rho_{m}(\mathbf{z})
 =\sum_{i=2}^{m}\alpha_{-i}
    \left(\ln(\mathbf{z}-\mathbf{w}_{i})
             -\ln(\mathbf{z}-\mathbf{w}_{-i})\right)
   +\rho(\mathbf{z})~.
\end{equation}
One can find that the correlation function (\ref{eq:DXminus})
can be expressed as a sum of the products of the connected
ones like
\begin{equation}
\left.\prod_{i=2}^{m}
     \left(2i\partial_{\alpha_{-i}}D_{\mathbf{w}_{i}}\right)
     \left\langle D\mathcal{X}^{-}(\mathbf{w}_{1})
     \right\rangle _{\rho_{m}}
\right|_{\alpha_{-i}=\theta_{i}=\theta_{1}=\Theta_{r}=0}\ ,
\label{eq:connected}
\end{equation}
with $m \leq n$, as the correlation functions are expressed
in terms of the connected ones in the usual field theory. 
%
%We would like to prove that the correlation function (\ref{eq:DXminus})
%has no singularities of the form (\ref{eq:correlation-psi}) satisfying
%above mentioned conditions, inductively with respect to $n$. Since
%we set all the $\Theta_{r}$ to be zero, there are no constant Grassmann
%odd parameters. Therefore, in the case of odd $n$, particularly the
%$n=1$ case, eq.(\ref{eq:DXminus}) vanishes and has no singularities.
%Let us prove the claim for $n=m(\geq2)$, assuming that it is true
%for $n\leq m-1$. Because of the assumption, 
It is therefore sufficient to prove
that the connected correlation function (\ref{eq:connected})
has no singularities of the form (\ref{eq:correlation-psi})
satisfying the conditions mentioned 
below eq.(\ref{eq:correlation-psi}).

The explicit form of 
$\langle D\mathcal{X}^{-}(\mathbf{w}_{1})\rangle_{\rho_{m}}$
can be obtained from eq.(4.1) of Ref.~\cite{Baba:2009fi} by replacing
$\rho$ with $\rho_{m}$. The super Mandelstam mapping $\rho_{m}$
possesses $N+2m-4$ interaction points. 
In the limit $\alpha_{-i}\rightarrow0$ for all $i$, 
$2m-2$ of them tend to $\mathbf{w}_{i},\mathbf{w}_{-i}$
$(i=2,\ldots,m)$ and the rest tend to the interaction points of $\rho$.
Let $\tilde{\mathbf{z}}_{I}^{[m]}$ denote the interaction point which
goes to the interaction point $\tilde{\mathbf{z}}_{I}$ of $\rho$
$(I=1,\ldots,N-2)$, 
%and $\tilde{\mathbf{z}}^{[m]}_{I^{(i)}}$,
%        $\tilde{\mathbf{z}}^{[m]}_{I^{(-i)}}$
%the interaction points
%which go to $\mathbf{w}_{i}, \mathbf{w}_{-i}$
%$(i=2,\ldots,m)$,
%respectively, 
in the limit.

Let us consider the terms in eq.(\ref{eq:connected}) that have poles
at $w_{1}=z_{I_{1}}$, which are relevant for us. These terms originate
from the terms in 
$\langle D\mathcal{X}^{-}(\mathbf{w}_{1})\rangle_{\rho_{m}}$
that have poles at $\mathbf{w}_{1}=\tilde{\mathbf{z}}_{I_{1}}^{[m]}$.
The residues of such poles are rational functions of 
$D\rho_{m}(\tilde{\mathbf{z}}_{I_{1}}^{[m]})$,
$\partial^{2}\rho_{m}(\tilde{\mathbf{z}}_{I_{1}}^{[m]})$ and higher
covariant derivatives of $\rho_{m}(\mathbf{z})$ at 
$\mathbf{z}=\tilde{\mathbf{z}}_{I_{1}}^{[m]}$,
with only powers of $\partial^{2}\rho_{m}$ in the denominators. (See
eqs.(4.1) and (B.3) in Ref.~\cite{Baba:2009fi}.) 
Let us apply 
$\prod_{i=2}^{m}\left(\partial_{\alpha_{-i}}D_{\mathbf{w}_{i}}\right)$
to such terms. The differentiation of $\tilde{\mathbf{z}}_{I_{1}}^{[m]}$
can be expressed by a rational function of the terms of the form 
$\left(\prod\partial_{\alpha_{-i}}\right)
  D^{l}\rho_{m}(\tilde{\mathbf{z}}_{I_{1}}^{[m]})$ with $l\geq 1$.
Therefore the results can be given by the terms of the form 
\begin{eqnarray}
 &  & \left(
\frac{1}{(\mathbf{w}_{1}-\tilde{\mathbf{z}}_{I_{1}}^{[m]})^{k}
        }~\mathrm{or}~\frac{\theta_{1}-\tilde{\theta}_{I_{1}}^{[m]}}
                           {(\mathbf{w}_{1}-\tilde{\mathbf{z}}_{I_{1}}^{[m]}
                             )^{k}}
\right)
\nonumber \\
 &  & \qquad\times
  \left(\mathrm{rational~function~of~}
        \left(\prod\partial_{\alpha_{-i}}\right)
                   D^{l}\rho_{m}(\tilde{\mathbf{z}}_{I_{1}}^{[m]})
  \right)~,
\end{eqnarray}
 with only powers of $\partial^{2}\rho_{m}$ in the denominators.
By taking $\alpha_{-i}=\theta_{i}=\theta_{1}=\Theta_{r}=0$, such
terms can have singularities only at $w_{i}=z_{I_{1}}$ as a function
of $w_{i}$. 
Namely, the correlation function of $\psi^{-}$ can have
singularities of the form (\ref{eq:correlation-psi}), but there
should be $i$ $\left(i\ne1\right)$ such that $z_{I_{i}}=z_{I_{1}}$.
Since such singularities do not satisfy the conditions
mentioned below eq.(\ref{eq:correlation-psi}),
they cannot contribute to 
the correlation function (\ref{eq:appidentity}).
Thus we have shown that eq.(\ref{eq:appidentity3}) holds.

\bibliographystyle{utphys}
\bibliography{SFTMar25_10}

\end{document}